\documentclass{IEEEtran4PSCC}

\ifCLASSINFOpdf
   \usepackage[pdftex]{graphicx}
\else
   \usepackage[dvips]{graphicx}
\fi

\usepackage{acronym}
\usepackage{cite}
\usepackage{xcolor}
\usepackage{multirow}
\usepackage[cmex10]{amsmath}
\usepackage{tabularx} 
\usepackage{array}   
\usepackage{makecell}
\usepackage{verbatim} 

\usepackage[colorinlistoftodos]{todonotes}

\newcolumntype{Y}{>{\centering\arraybackslash}X}

\makeatletter
\let\old@ps@headings\ps@headings
\let\old@ps@IEEEtitlepagestyle\ps@IEEEtitlepagestyle
\def\psccfooter#1{%
    \def\ps@headings{%
        \old@ps@headings%
        \def\@oddfoot{\strut\hfill#1\hfill\strut}%
        \def\@evenfoot{\strut\hfill#1\hfill\strut}%
    }%
    \def\ps@IEEEtitlepagestyle{%
        \old@ps@IEEEtitlepagestyle%
        \def\@oddfoot{\strut\hfill#1\hfill\strut}%
        \def\@evenfoot{\strut\hfill#1\hfill\strut}%
    }%
    \ps@headings%
}
\makeatother

\begin{document}

\title{TSO–DSO Coordination for Flexibility\\ Management Across Voltage Levels}

\author{\IEEEauthorblockN{Gustavo Valverde\IEEEauthorrefmark{1},
Anibal Sanjab\IEEEauthorrefmark{2},
Florin Capitanescu\IEEEauthorrefmark{3}, 
Christian Rehtanz\IEEEauthorrefmark{4},\\
Gianluigi Migliavacca\IEEEauthorrefmark{5},
Zhengshuo Li\IEEEauthorrefmark{6}, and 
Innocent Kamwa\IEEEauthorrefmark{7} 
}
\IEEEauthorblockA{\IEEEauthorrefmark{1} Power Systems Laboratory, 
ETH Zurich,
Switzerland}
\IEEEauthorblockA{\IEEEauthorrefmark{2} Flemish Institute for
Technological Research (VITO) and EnergyVille, Belgium}
\IEEEauthorblockA{\IEEEauthorrefmark{3} Luxembourg Institute of Science and Technology, Luxembourg}
\IEEEauthorblockA{\IEEEauthorrefmark{4}Technical University Dortmund, Germany}
\IEEEauthorblockA{\IEEEauthorrefmark{5} Ricerca sul Sistema Energetico (RSE), Italy}
\IEEEauthorblockA{\IEEEauthorrefmark{6} Shandong University, China}
\IEEEauthorblockA{\IEEEauthorrefmark{7} Université Laval, Canada}
}

\maketitle
\begin{abstract}
 Several sources of flexibility in transmission and, especially, distribution networks are being unlocked by advances in information and communication technologies, aggregators, and new flexibility markets. However, maximizing benefits for both transmission and distribution system operators in a coordinated way requires new algorithms, modeling tools, and modernization of regulatory frameworks. Such approaches must account for uncertainties, the physical and operational constraints of flexibility providers and the grid itself, constraints on information exchange, and scalability, including computational requirements and time constraints. 
 
 Given the diverse contexts and jurisdictions around the world, there is no single recipe for achieving coordination, but important trends and shared challenges are emerging. This paper surveys the complexities of coordination from technical, market, and technological perspectives, and outlines current practices, proposed approaches, and future research directions to effectively manage, coordinate, model, and leverage flexibility across voltage levels.    
\end{abstract}


\thanksto{\noindent Invited survey paper to the 24th Power Systems Computation Conference (PSCC 2026). Anibal Sanjab has been supported by the U2Demo project, funded under the European Union's Horizon Europe Programme, grant agreement no. 101160684. Florin Capitanescu acknowledges the support of Luxembourg National Research Fund (FNR) in the frame of the project TESTIFY (C21/SR/15762760).}

\section*{List of Acronyms}

\begin{acronym}[CMPLDWG]
\acro{AC}{Alternating Current}
\acro{A-DERMS}{Aggregator DERMS}
\acro{ADMM}{Alternating Direction Method of Multipliers}
\acro{ADMS}{Advanced Distribution Management System}
\acro{ADN}{Active Distribution Network}
\acro{AI}{Artificial Intelligence}
\acro{AMI}{Advanced Metering Infrastructure}
\acro{ANN}{Artificial Neural Network}
\acro{AS}{Ancillary Service}
\acro{BA}{Balancing Authority}
\acro{BESS}{Battery Energy Storage System}
\acro{CMPLDWG}{Composite Load Model with DG} 
\acro{DA}{Day-ahead}
\acro{DC}{Direct Current}
\acro{DER}{Distributed Energy Resource}
\acro{DERMS}{DER Management System}
\acro{DG}{Distributed Generation}
\acro{DN}{Distribution Network}
\acro{DNO}{Distribution Network Operator}
\acro{DP}{Distribution Provider}
\acro{DSO}{Distribution System Operator}
\acro{EB-GL}{Electricity Balancing Guidelines}
\acro{EMS}{Energy Management System}
\acro{EV}{Electric Vehicle}
\acro{FERC}{Federal Energy Regulatory Commission}
\acro{FR}{Flexibility Region} 
\acro{FSP}{Flexibility Service Provider}
\acro{HP}{Heat Pump}
\acro{HV}{High Voltage}
\acro{ICT}{Information and Communications Technology}
\acro{ISO}{Independent System Operator}
\acro{IT}{Information Technology}
\acro{LFM}{Local Flexibility Market}
\acro{LP}{Linear Programming}
\acro{LSE}{Load-Serving Entity}
\acro{LSTM}{Long Short-Term Memory}
\acro{LV}{Low Voltage}
\acro{ML}{Machine Learning}
\acro{MO}{Market Operator}
\acro{MV}{Medium Voltage}
\acro{NAN}{Neighborhood Area Network}
\acro{NERC}{North American Electric Reliability Corp.}
\acro{NLP}{Non-Linear Programming}
\acro{OE}{Operating Envelope}
\acro{OFO}{Online Feedback Optimization}
\acro{OLTC}{On-Load Tap Changer}
\acro{OPF}{Optimal Power Flow}
\acro{OT}{Operational Technology}
\acro{PLC}{Power Line Communication}
\acro{PMU}{Phasor Measurement Unit}
\acro{PV}{Photovoltaic}
\acro{RES}{Renewable Energy Source} 
\acro{RSF}{Residual Supply Function}
\acro{RTO}{Regional Transmission Organization}
\acro{SCADA}{Supervisory Control and Data Acquisition} 
\acro{SLMR}{Synthesis Load Model with Renewables} 
\acro{SO}{System Operator}
\acro{SV}{Shapley Value} 
\acro{TD}[T\&D]{Transmission and Distribution}
\acro{TN}{Transmission Network}
\acro{TNO}{Transmission Network Operator}
\acro{TOP}{Transmission Operator}
\acro{TSO}{Transmission System Operator}
\acro{U-DERMS}{Utility DERMS}
\acro{VFD}{Variable Frequency Drive}
\acro{VPP}{Virtual Power Plant}
\acro{WAN}{Wide Area Network}
\acro{WECC}{Western Electricity Coordinating Council}
\end{acronym}

\section{Introduction}

\subsection{Historical Evolution of Transmission and Distribution Systems Interactions}

For most of the twentieth century, electric power systems were conceived, planned, and operated under a vertically hierarchical paradigm in which transmission and distribution networks fulfilled clearly delineated and mostly independent functions. High-voltage transmission systems -- operated by \acp{TNO} or, later, \acp{TSO} -- were engineered to ensure bulk power transfer, system-wide balancing, and frequency regulation across large geographic areas. In contrast, \acp{DN} -- owned and managed by \acp{DNO} -- were designed as predominantly passive infrastructures, responsible for delivering electrical energy from transmission substations to end users according to conservative planning and reliability criteria. Under this paradigm, the transmission-distribution interface was characterized by limited operational interaction, high predictability of power flows, and a unidirectional exchange of energy from the bulk system to the distribution level.

This separation was institutionally reinforced by regulatory and market designs. In Europe, the unbundling reforms of the 1990s and early 2000s formalized the distinction between \ac{TSO} and \ac{DNO} responsibilities, while preserving a predominantly passive role for \acp{DN} \cite{ENTSOE_2015}. In North America, many utilities remained vertically integrated, owning and operating \acp{DN} and, in some cases, transmission assets, while \acp{ISO} and \acp{RTO} later assumed bulk-system operational and market functions in parts of the system. Despite these institutional differences, \acp{DN} were treated as an electrically subordinate
layer, largely invisible to real-time system operations. As a result, planning, security assessment, and \acp{AS} procurement were executed almost exclusively at the transmission level, with distribution systems represented through aggregated and static equivalents \cite{CIGRE_TB845_2021}.

The European and North American frameworks partition power system operational responsibility along fundamentally different axes. EU electricity law distinguishes legally defined network operators: the \ac{TSO}, responsible for transmission-system operation and system balancing, and the \ac{DSO}, responsible for distribution-system operation downstream of the TSO–DSO interface, typically at medium- and low-voltage levels. These roles are subject to unbundling requirements separating network operation from competitive generation and supply activities; where a \ac{DSO} forms part of a vertically integrated undertaking, EU rules require legal, functional, and accounting separation, subject to exemptions for smaller \acp{DSO}.

By contrast, \ac{NERC} assigns reliability obligations through functional entity roles rather than through legally distinct operator categories tied to voltage level or ownership structure. Consequently, a single legal entity may simultaneously perform and be registered for several reliability functions, such as \ac{TOP}, \ac{BA}, and \ac{DP}. While \ac{NERC} \ac{TOP} and \ac{BA} roles map closely to the network and balancing mandates of a European \ac{TSO}, the \ac{NERC} \ac{DP} carries a narrower, reliability-only scope compared to the \ac{DSO}’s broader European mandate for market facilitation and resource procurement. Furthermore, the \ac{NERC}-defined \ac{LSE} has no direct European counterpart, where retail supply is a commercial function structurally decoupled from network operation.

Over the past two decades, these long-standing paradigms have been progressively disrupted by the rapid diffusion of \acp{DER} -- including inverter-based renewable generation, distributed storage, flexible demand, \acp{EV}, and other digitally controlled loads -- connected predominantly at the distribution level \cite{IEEEStd2030}. As documented across multiple jurisdictions, \acp{DN} are no longer passive sinks of power flows, but active interfaces where variability, flexibility, and controllability increasingly originate. Distribution-level actions can now materially influence transmission congestion, voltage stability, frequency response, and inter-area dynamics. 

This evolution has forced a conceptual shift from the traditional transmission and distribution network operator model -- focused primarily on asset ownership and infrastructure -- to the more operationally oriented \ac{TSO}–\ac{DSO} framework, in which \acp{SO} must coordinate in real time and across planning horizons to preserve security, efficiency, and market integrity \cite{Coujard2022}. The distinction is not merely semantic. While \acp{DNO} historically optimized local reliability and capital expenditure, \acp{DSO} are increasingly expected to perform system-level functions such as congestion management, voltage control, flexibility procurement, energy balancing, active network operation, and data orchestration \cite{TSO_DSO_ASM_2019}. Consequently, the transmission–distribution boundary has evolved into a cyber-physical interface where operational authority, control actions, and market outcomes intersect \cite{CaiMengmeng2025}.

European policy institutions recognized this shift early. ENTSO-E explicitly framed distribution-connected resources as essential contributors to system-wide services and called for revised operational, market, and data-handling arrangements between \acp{TSO} and \acp{DSO}, while emphasizing that no single coordination model can be universally applied given national diversity in grid structures and regulatory traditions~\cite{ENTSOE_2015}. Building on this recognition, research initiatives such as \cite{GERARD2018, Papavasiliou2018,  MarcoRossi2020,SanjabAnibal2022, MarquesLuciana2023, AliHajebrahimi2024} systematically evaluated alternative coordination schemes -- ranging from centralized \ac{TSO}-led markets to shared or common \ac{TSO}–\ac{DSO} flexibility platforms -- using detailed transmission–distribution co-simulation and cost–benefit analysis. These studies demonstrated that uncoordinated activation of distribution-connected flexibility leads to inefficient outcomes and potential security violations, whereas coordinated architectures consistently outperform siloed approaches across multiple national scenarios \cite{FCapitanescu2018,SanjabAnibal2022,MarquesLuciana2023,AliHajebrahimi2024}. Their findings served to shape European approaches regarding coordinated flexibility markets, priority rules, and shared operational responsibilities.

Building on these policy and research developments, balancing and \ac{AS} markets have also continued to evolve, as illustrated by the EU’s \ac{EB-GL} and the establishment of cross-border balancing platforms such as MARI~\cite{MARIWebpage}, PICASSO~\cite{PICASSOWebpage}, and TERRE~\cite{TERREWebpage}. In parallel, increasingly stressed grid conditions -- driven by electrification and the large-scale integration of \acp{DER} at the distribution level -- are also raising flexibility needs for \acp{DSO}. Accordingly, several \ac{LFM} initiatives are emerging (e.g., as mapped in \cite{VITO_2025}), alongside regulatory developments such as the forthcoming EU network code on demand response~\cite{NCDRWebpage}. The combination of rising flexibility needs at both transmission and distribution levels, and the growing availability of flexible resources across these grid layers, further motivates TSO–DSO coordination in flexibility procurement and activation, with the aim of improving procurement efficiency while ensuring that flexibility actions remain grid-safe for all networks involved.

In the United States, distribution utilities largely remain vertically integrated monopolies regulated at the state level, while bulk power system operation is performed by \acp{ISO}, \acp{RTO}, or \acp{BA} \cite{DebraLew2017}. The adoption of \ac{FERC} Order No. 2222 marked a structural turning point by mandating that \ac{DER} aggregations be permitted to participate directly in wholesale markets~\cite{FERC_Order_2222_2020}. This regulatory intervention effectively forced the creation of formal transmission-distribution coordination mechanisms -- particularly for real-time operations, deliverability assessment, override authority, and settlement -- without redefining distribution utilities as independent \acp{DSO} \cite{EllaZhou2222}. As articulated in recent U.S. Department of Energy frameworks \cite{DOE_TSO_DSOAgg2024}, scalable market and operational coordination platforms are emerging as pragmatic solutions to manage the dual use of \acp{DER} for wholesale and distribution services while preserving reliability and regulatory jurisdictional boundaries.

In Canada, and particularly in Ontario, the transition toward active distribution operation has been more incremental and explicitly policy-driven. The Ontario Energy Board’s 2025 Discussion Paper frames the \ac{DSO} not as a replacement for the \ac{DNO}, but as a graduated enhancement of distribution capabilities, progressively enabling planning, operational, and market functions as DER penetration increases. In this respect, the Ontario approach has emphasized functional evolution within existing utilities, with careful attention to legislative permissibility, cost recovery, and consumer protection. Complementary work by the Independent Electricity System Operator further highlights the practical challenges of transmission–distribution coordination in a hybrid \ac{ISO}–utility environment \cite{IESOTD2021}.

Québec represents a distinctive yet instructive variant of this evolution where the legal and regulatory framework for the electricity sector is overseen by the \textit{Régie de l’énergie}, an independent economic tribunal that regulates tariffs, distribution conditions, and reliability requirements for Hydro-Québec and other utilities. Historically, both transmission and distribution functions have been operated within a vertically integrated public utility structure, effectively internalizing what elsewhere would constitute explicit \ac{TSO}–\ac{DSO} interfaces. This model enabled system-wide optimization based on abundant hydroelectric resources, high inertia, and predictable load profiles, long justifying the treatment of \acp{DN} as passive extensions of the bulk system. However, accelerated electrification, growing penetration of distribution-connected \acp{DER}, large flexible loads, and increasing interprovincial and cross-border exchanges are introducing operating conditions in which distribution-level actions can materially affect transmission security and dynamic performance.

Compared to Ontario’s regulator-led, incremental development of explicit \ac{DSO} capabilities and the United States’ market-driven, \ac{ISO}-centric coordination under \ac{FERC} Order No. 2222, Québec occupies an intermediate position. Here, long-standing vertical integration has delayed formal \ac{TSO}–\ac{DSO} separation while simultaneously enabling earlier deployment of internally coordinated, system-wide operational mechanisms that can naturally evolve into stability-aware \ac{DSO} functions without wholesale institutional restructuring. However, Hydro-Québec’s mandatory annual reporting to the \textit{Régie de l’énergie} on distribution activities, including service quality and connection performance, highlights active regulatory oversight of distribution operations that can inform the evolution toward broader \ac{DSO}-like capabilities.

China primarily adopts a vertically integrated monopoly system, where the \ac{HV}, \ac{MV}, and \ac{LV} grids are planned, constructed, and operated by unified grid enterprises. While the organizational structure is centralized, functional responsibilities are divided into layers: the \ac{TN} is controlled by the transmission control center through the \ac{EMS} \cite{MikeZhou2018}, and the \ac{DN} is managed by the distribution control center using the \ac{ADMS} \cite{WenchuanWu2022}. This management model aims to ensure the overall coordination of the power system. 

The coordination between transmission and distribution in China is mainly based on vertical control instructions within a unified scheduling system, rather than through cooperation between separate organizations such as the \ac{TSO}, \ac{DSO}, and \acp{MO}. This structure facilitates resource allocation, rapid emergency response, and low coordination costs. However, it may also lead to insufficient market competition, thereby limiting the potential to fully exploit and utilize flexible resources. 

With the large-scale integration of \acp{DER}, some regions in China are facing new challenges, such as reverse power flows and voltage violations in \acp{DN} \cite{ChenLuo2021}. In response, China is advancing power-sector reforms and introducing market-based competition mechanisms to overcome the limitations of the current management model.
In the future, the coordination between transmission and distribution in China may evolve from an internal administrative model to a multi-agent model based on market signals and information sharing.

\subsection{Paper Scope and Thrust}

Despite these differing institutional pathways, several common technical challenges emerge across jurisdictions. These include the need for enhanced observability and state awareness across the transmission–distribution boundary; explicit coordination of flexibility procurement, activation, and priority rules to not only avoid conflicting control actions but also to capitalize on the value stacking potential of flexibility -- whereby flexibility activations can simultaneously meet the needs of multiple \acp{SO} (at transmission and distribution levels); consistent market coupling and settlement mechanisms that incentivize flexibility deliver and prevent double compensation; clear definition of operational authority and fail-safe override procedures under contingencies; and coordinated planning processes that align transmission reinforcement with distribution-level non-wires alternatives. Evidence from European, U.S., and Australian coordination studies demonstrates that these challenges cannot be addressed through market design alone, but require integrated architectures combining physical network models, aggregation layers, and coordination platforms capable of operating coherently across multiple time scales \cite{Givisiez2020}. 

Against this backdrop, the present paper focuses on the technical challenges and coordination models for \ac{TSO}–\ac{DSO} coordination in the context of flexibility markets, network planning and system operation, architectural options for information exchange and control, and the implications of differing regulatory and institutional contexts. Rather than advocating a single coordination model, the paper adopts a comparative and architecture-centric perspective, identifying invariant technical requirements that persist across jurisdictions despite differences in governance and market structures. 

The paper and the surveyed coordination approaches are explored with regard to operational planning (e.g., day-ahead), expansion planning where relevant, operation, and TSO-DSO coordinated flexibility market procurement, which can run at different time scales from long-term to short-term to near real-time. Within operation, the focus is mainly on secondary and tertiary open-loop control. 

In view of the five operating states of a power system~\cite{Fink1978}, the paper and the surveyed coordination approaches mainly explore normal and alert states, while more extreme states (i.e., emergency, in extremis, and restorative) that require typically fast reactions are to be researched. Accordingly, fast phenomena (e.g., electromagnetic transients, protection schemes) as well as defense planning and restoration are outside the scope of this survey. 

This paper does not intend to perform a comprehensive review of the state of the art in \ac{TSO}--\ac{DSO} coordination, but rather to explore different dimensions of coordination—including planning, operations, and flexibility markets—while highlighting the key theoretical and practical challenges and opportunities inherent in each.

By grounding the analysis in established large-scale studies, and emerging technical standards \cite{IEEEStd15472,EllaZhou2222}, as well as in large-scale pilot activities (e.g, ~\cite{OneNetWebpage}), the paper aims to bridge the gap between conceptual coordination models and their practical implementation in increasingly DER-dominated, distribution-active power systems.

\section{Flexibility of Distribution Systems as the enabler of TSO-DSO coordination}

\subsection{Sources of Flexibility and Technical Constraints}

The need for flexibility is increasing for both \acp{DSO} (e.g., for congestion management and voltage control) and \acp{TSO} (e.g., for balancing services, congestion management, and voltage control).

This work understands flexibility as the ability of an asset (e.g. \ac{DER}) to adjust its active/reactive power (or voltage magnitude) within its physical and operational limits, enabling it to respond to system needs and provide grid services. The key characteristics of flexibility are power, energy (including the capacity to maintain a given power setpoint), ramp-up and ramp-down rates, and reliability of service. Note that in power systems, there are other, network-level assets that allow \acp{SO} operational flexibility by adjusting network topology and operating conditions through devices such as switches, phase-shifting transformers, \ac{OLTC} transformers, and DC links, thereby altering active or reactive power flows~\cite{VanHertem2005}. 

At the \ac{TN} level, typical sources of flexibility include conventional and renewable generation units with fast-ramping and regulation capabilities, large-scale energy storage systems, cross-border interconnections, and assets that can adjust reactive power at substations (e.g., slow capacitor/reactor banks and fast SVCs or STATCOMs). Flexibility may also be provided by the dynamic rating of overhead power lines~\cite{TengFei2018}, underground cables, and transformers, an option that is increasingly being considered. Finally, controllable large loads directly connected to the \acp{TN} can offer flexibility services.

As the power system decentralizes, additional diverse sources of flexibility are emerging in the \ac{DN}, among them: conventional and inverter-based \ac{DG} units, utility-scale and behind-the-meter \acp{BESS}, as well as flexible loads including \acp{HP}, air conditioners, cooling systems, fans, electric water heaters, and \acp{EV}. Nevertheless, the efficient use of existing network-side flexibility resources should be considered alongside DER-based flexibility. These include \acp{OLTC}, capacitor/reactor banks, dynamic rating, and topology control. Key options include optimal radial network reconfiguration \cite{NEBULONI2025}, and in the near future, soft open points \cite{ShuhanLi2024,JiangXun2025} and MVDC links \cite{Coffey2021}. 

Due to the size of utility-scale \acp{DER}, they are mostly monitored by the \ac{DSO}. Grid codes should require that \acp{DER} provide dynamic voltage and frequency regulation, provided they remain within the ride-through zone and within their capability curves~\cite{Hatziargyriou2017BulkSystemDER}. For voltage support, inverters may operate in \textit{Q-priority} mode, curtailing active power to prioritize reactive power when the apparent-power limit is reached. Alternatively, and to avoid active power curtailment, some inverter-based \acp{DER} can inject or consume reactive power beyond their capability curves for short periods with little or no impact on inverter lifetime \cite{IEEEStd15472}. 

Regarding frequency regulation, the majority of \ac{DG}-type \acp{DER} operate at their maximum available power, only able to provide downward support during over-frequency conditions unless they hold headroom in reserve for upward support during under-frequency conditions, while \acp{BESS} may provide active-power frequency control for both downward and upward support. Although \acp{DERMS} could enable \acp{DER} to provide upward support when \acp{DER} are curtailed, this would need coordination between the \ac{TSO} and the \ac{DSO} to confirm that the frequency-related response does not lead to any reliability issues that led to the original \ac{DER} curtailment~\cite{IEEEStd15472}. 

Small-scale \acp{DER}, such as rooftop \ac{PV} systems and behind-the-meter \acp{BESS}, may respond to voltage and frequency events using local measurements but can also participate in \ac{TN} services when coordinated by aggregators or \acp{VPP} \cite{GValverde2019,Karagiannopoulos2021, Srivastava2025}. Additionally, the four-quadrant capability of \acp{BESS} makes them a valuable source of flexibility for congestion management, peak shaving, and voltage/frequency regulation; however, their scheduling and dispatch must be optimized to balance owner objectives with the provision of grid services. 

Only a few works have investigated the coordination of utility-scale and behind-the-meter \acp{DER} for the same service provision. Some efforts include power factor correction at the \ac{TSO}-\ac{DSO} interface \cite{GValverde2019}, regulation requests from the transmission system~\cite{DallAnese2018}, and system restoration \cite{LiuWeijia2021}. These are coordination schemes that require further development but will be enabled by \ac{DERMS}. 

On the demand side, thermostatically controlled loads such as water heaters, \acp{HP}, and air conditioners offer opportunities for flexibility given their thermal inertia~\cite{Buechler2024}. However, coordination is needed to avoid load synchronization and rebound effects once a demand-reduction service ends \cite{Agah2024}. Modern inverter-driven \acp{HP} can adjust their active power consumption for grid services, while older \ac{HP} units under ON--OFF control can still be blocked/unblocked, subject to flexibility constraints, e.g., minimum ON and OFF times and a maximum number of starts per day to protect the \ac{HP}'s lifespan.   

Among the flexible loads, the \acp{EV} are probably the easiest to monitor and control via application programming interfaces, allowing the retrieval of battery state of charge, charging status, and remote start and stop of charging sessions. Nevertheless, as non-fixed assets, there remains some uncertainty about when the vehicles will be available for service provision, even though the plug-in sessions are mostly predictable. Additionally, after charging-session interruptions, there may be restrictions on subsequent charging due to simultaneous charging by nearby \acp{EV}.  

The participation of \acp{DER} in services (balancing, voltage regulation, congestion management, etc.) should not cause voltage-limit violations, equipment overloading, excessive power losses, or phase imbalance, and must not lead to excessive operation of voltage-regulation equipment such as substation \acp{OLTC}, line voltage regulators, or switched capacitor banks. Therefore, full exploitation of \acp{DER} flexibility is constrained by the local grid. The limitations depend on the location and concentration of \acp{DER}. For example, flexible devices closer to the distribution transformer are less voltage-related restricted than those at the end of the \ac{LV} feeders, and dispersed \acp{DER} face fewer restrictions than \acp{DER} concentrated in \ac{DER}-dense feeder sections.

Grid characteristics also affect the amount of flexibility that can be leveraged. In North America, \ac{MV} feeders leaving the substation are generally long and ramify into three-phase and single-phase laterals. Along these feeders and laterals, hundreds of small-capacity distribution transformers are connected to supply customers directly or through \ac{LV} (secondary) systems, which are generally short. In residential areas, these secondary systems are single-phase, three-wire circuits, radially operated at 120/240~V, feeding tens of customers. Here, phase imbalance at the MV level and potential overloading of distribution transformers could limit the provision of flexibility from \ac{LV}-connected \acp{DER}. Additionally, for comparable feeder design and loading, voltage limits in the \ac{MV} level may be exceeded more often in 15 kV-class circuits than in 25- or 35-kV systems, due to higher currents and thus larger voltage drops. 

In Europe and other regions, \ac{MV} feeders are often shorter than in North America and supply three-phase distribution transformers of several hundred kVA that feed three-phase, four-wire \ac{LV} systems at 230/400~V. In this context, voltage and thermal constraints on the \ac{LV} network can become limiting, particularly in areas with high load density and \acp{DER}. 

Finally, some \ac{DN} characteristics do not merely limit how much \ac{DER} flexibility can be used; they may also determine which type of flexibility is most valuable or even required locally. For example, \acp{DN} with long underground cables may result in excessive reactive power being exported to higher voltage levels~\cite{AnnaShuvarina2024}. This may force \acp{DER} and other assets to primarily provide reactive power compensation or local voltage support, preventing them from offering other services.

\subsection{Feasible Regions of Aggregated \ac{DN} Flexibility at the Interface with the \ac{TN}}
\label{sec:feasible_regions}

Chronologically, the earliest approaches to underpin TSO-DSO coordination emerged in 2018 \cite{JoaoSilva2018,FCapitanescu2018,Mayorga2018,Contreras2018}. Based on various \ac{OPF} formulations, they have focused on approximating the boundary of the feasible region of an \ac{ADN} in the space of active (P) and reactive (Q) power flows at the transmission-distribution interface (i.e. physical substation) for a single point in time. Such a region (also known as a PQ chart, flexibility region, or map) characterizes the \ac{DER}'s aggregated active/reactive power flexibility in the \ac{ADN} that meets its operational constraints. It enables a coordination mechanism that preserves confidentiality while ensuring technical feasibility. 

The computation of the \ac{FR} typically aims to maximize the power range at the interface, subject to a power-flow formulation (equality constraints) and imposed voltage and power flow limits (inequality constraints). In this regard, the use of different power flow models has been explored in the literature, from single-phase to three-phase imbalance calculations, to linear, second-order cone programming, or non-linear non-convex \ac{AC} formulations. The effectiveness, degree of inaccuracy, and sub-optimality of each one of these methods vary widely and have been extensively analyzed in the literature. Although linearized power flow models to approximate the PQ chart offer computational advantages, the approximation errors may affect operational decisions in practical \ac{TSO}-\ac{DSO} coordination, see Section~\ref{sec:FR_integration_market}. 

Recent works indicate that the \ac{FR} under the full \ac{AC} power flow model may be non-convex \cite{JoSilva2018,Prionistis2025,NNazir2022,DavidPozo2025}, discontinuous \cite{Prionistis2025,DavidPozo2025} including exhibiting holes \cite{NNazir2022}, and it changes with the discrete steps of the \ac{OLTC} in step-down transformers at the substation \cite{JoSilva2018} and with network reconfigurations \cite{FlorinCapitanescu2024}. These issues, which academia will continue to address from a theoretical perspective, are common to many other problems in power systems and should not hinder the development of practical approaches (i.e. targeting simplicity, accuracy, scalability and usability in a market framework) to support \ac{TSO}-\ac{DSO} coordination in real-life applications. 

The review so far has focused on open-loop preventive or corrective control, while further work should also examine potential closed-loop coordination schemes such as \ac{OFO}~\cite{PICALLO2020,KLEINHELMKAMP2024,ZHAN2025111468}. These methods enable near-optimal operation while enforcing grid constraints with minimal model information of the nonlinear system~\cite{ORTMANN2026}.

\subsection{Cost of Flexibility of Aggregated \ac{DN} Flexibility at the interface with the \ac{TN}}

Despite extensive research efforts, the \ac{ADN} technical feasible regions are insufficient for \ac{TSO}-\ac{DSO} coordination as they do not map analytically any point in the region to a cost \cite{FlorinCapitanescu2024}. In other words, they do not respond to the question: how much would it cost to the \ac{TSO} to shift the active or reactive power flow performed by the \ac{DSO} at their interface? 
Furthermore, the relationship between the cost and the amount of active/reactive power flow flexibility must be computed fast and expressed as tractable constraints, so as to be seamlessly embedded in \ac{TSO} tools for procuring flexibility at the least cost, including from \acp{ADN}, without compromising computation time~\cite{FlorinCapitanescu2024}. Related aggregation frameworks have considered the aggregation of \ac{DER} cost functions within \acp{VPP}, either using linearized distribution power flow models~\cite{Zhongkai2020} or neglecting network constraints altogether~\cite{MingQu2020}. However, these approaches do not fully address the problem of deriving tractable cost-flexibility mappings at the \ac{TSO}-\ac{DSO} interface while accounting for binding \ac{DN} constraints.

Deriving cost curves of active/reactive power flexibility at the \ac{TSO}-\ac{DSO} interface based on \acp{FR} is challenging as the cost curves cannot be obtained in analytical form due to the nonlinearity of \ac{ADN} operation constraints, but especially binding (voltage and thermal) constraints \cite{JoSilva2018}. Brute-force computation of the cost in each PQ point of the region has been attempted \cite{MarcelSarstedt2022,AChurkin2024}; yet it is not scalable (as it relies on solving a huge number of \ac{AC} \ac{OPF} problems) or usable (as it does not derive specific simple mapping constraints). 

To circumvent these issues, from an operational perspective, a sub-optimal yet practical and computationally effective approach has been proposed for a single period \cite{FlorinCapitanescu2024}. It considers the ranges of active and reactive power separately, as they mitigate different issues (e.g., congestion vs. voltage) using different efficient control means. It develops piecewise linear cost curves for each of them by solving a series of tailored \ac{AC} \ac{OPF} problems. This approach has been further leveraged to consider coupling constraints (e.g. due to the presence of storage devices, flexible demand, and limit of \acp{OLTC}) for active power \cite{FlCapitanescu2025} and reactive power \cite{FlorCapitanescu2025} separately. These works have provided new insights and research directions to overcome some of the identified barriers (e.g., guaranteeing the feasibility of the proposed flexibility patterns across all time periods in a scalable way), which should spur further improvements. 

From a market integration perspective, dedicated research on computing price/quantity active power flexibility pairs at the \ac{TSO}-\ac{DSO} interface region was developed in~\cite{Ananduta2025_BidForwarding} and initially conceptualized in~\cite{Papavasiliou2018}. These works iteratively solve a linear \ac{OPF} problem, that models active power flow and thermal constraints only, at the distribution level considering an increment in the \ac{TSO}-\ac{DSO} interface active power flow. It generates a solution through which a \ac{DSO} can provide an increment volume of flexibility to the \ac{TSO}, which is conceptually resembling to \cite{FlorinCapitanescu2024}. The solution of this problem also returns the optimal price for offering this incremental volume of flexibility. Based on an iterative solution to this problem, each with an additional interface power increment, these works compute what are referred to as \acp{RSF}, which constitute step-wise price/quantity pairs of flexibility that the \ac{DSO} can provide to the \ac{TSO}. The \ac{TSO} can then choose one of these steps, which the \ac{DSO} can then readily activate via the \ac{OPF} solution corresponding to this activation step. The sub-optimality decreases with a smaller increment size, but in turn, increases the computational time. This tradeoff, in addition to regulatory challenges in terms of bid aggregation by a third market-neutral party (e.g., \ac{DSO} or independent \ac{MO}) poses practical challenges to the implementation of this approach. A similar approach has been proposed in \cite{YujiaLi2025}, but at the planning stage.   

In reality, the above game is more complex. It involves \acp{TSO}, \acp{DSO}, aggregators, suppliers, and market operators.  Depending on the organization, market design, and regulatory framework, contract definitions can vary significantly. For \acp{TSO} and \acp{DSO} whose sole mission is grid operations, interaction with grid users should occur through \acp{AS} to ensure stable and secure operation. Market clearing between players is not their responsibility.

\subsection{Aggregation and Control of DERs for Grid ASs}

Several schemes for controlling and aggregating \acp{DER} for grid \acp{AS} (e.g., voltage and frequency control) have been proposed recently. 

Voltage support through the control of active and reactive power exchange at the TSO-DSO interface~\cite{GValverde2019,Karagiannopoulos2021,Alizadeh2023}, and voltage-stability support during emergency conditions~\cite{LuisOspina2021,PRIONISTIS2022} are two researched aspects. Some of them use \ac{PMU} measurements at \ac{TN} buses to define system conditions and the need for reactive power provision from \acp{ADN}~\cite{FEscobar2022,Rousis2021}. Future schemes must leverage \ac{DER} participation optimally to avoid over- and under-responses, inefficiencies, and unnecessary \ac{DER} saturation at their operational limits (active/reactive power, or state of charge), as \acp{DER} from one substation may be more suitable to participate than those from other substations. However, planning voltage support in \ac{DA} under the worst realization of uncertainties cannot be well-posed and requires further thought~\cite{FloCapitan2024}.

For frequency regulation, previous works have proposed the use of \acp{ADN}, \acp{DER}, and \ac{EV} aggregations~\cite{Kontis2021,Miletic2024,SekyungHan2010,Wenzel2018}, and fast-frequency control with thousands of \acp{DER} \cite{Lundstrom2021_part1,Lundstrom2021_part2}.

Controlling large populations of \acp{DER} for \ac{DN} and \ac{TN} support poses several challenges. The first is the amount of information that can be transmitted and processed in real time, particularly for control schemes that rely on direct communication with thousands of \acp{DER} and flexible loads. The second one relates to the uncertainty of availability when the service is needed, i.e., how much and for how long \ac{DER} units can maintain it. The third one is the service type they will offer to the \ac{DSO} and the \ac{TSO}. For example,  voltage stability support to the \ac{TSO} and local voltage control to the \ac{DSO} \cite{AristidouPetros2015}, or frequency regulation provision to the \ac{TSO}, and voltage-control services to the local grid \cite{Karagiannopoulos2020,ZottiGiulia2020}.  

Conflicting requests between \acp{SO} will require coordination or the definition of priorities~\cite {ZottiGiulia2020}. Competition between \ac{TSO} and \acp{DSO} in procuring flexibility from \acp{DER} and flexible loads, as well as coordination mechanisms, is discussed in Section~\ref{sec:market_aspects}. The fourth challenge is the effect of such \ac{DER} services in the local grid. Participation of \acp{DER} for \ac{TN} support should not deteriorate the operation of the \ac{DN}, for example, creating over- or under-voltages, equipment overloading, or increasing phase unbalance. 

While previous academic works and technical reports have pointed out the feasibility of service provision~\cite{Hatziargyriou2017BulkSystemDER}, modeling simplifications of \acp{DN} may hide operational problems that would limit greater \ac{DER} and \ac{ADN} participation in grid services. For example, the use of balanced \ac{DN} representations despite many \acp{DER} and flexible loads being single-phase devices, whose participation in services may result in unacceptable phase-imbalance conditions. Another common assumption is using a single radial feeder to represent a \ac{DN} rather than multiple feeders connected to the same substation.

\subsection{Dynamic Equivalents of \acp{ADN}}

\begin{figure}[h!]
    \centering
    \includegraphics[width=0.45\textwidth]{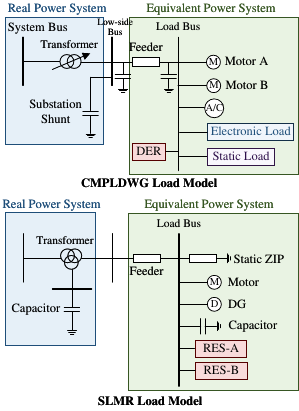}
    \caption{Equivalents of active distribution networks for stability studies.}
    \label{fig:Dyn_Eq_ADNs}
\end{figure}

\acp{ADN} with high levels of \acp{DER} will impact the steady-state and dynamic performance of power systems~\cite{Hatziargyriou2017BulkSystemDER}. For stability studies conducted by \acp{TSO}, a key challenge is how to represent the response of \acp{ADN} during disturbances in the \ac{TN}. This problem arises because \acp{TSO} lack access to detailed \ac{DN} models, as well as granularity regarding the number, location, and types of \acp{DER}. Compounding this issue, \acp{DSO} have historically not required dynamic models, as these analyses were traditionally confined to the transmission level.

Industry and academia have addressed this by using composite load models with aggregated representations of \ac{DER} populations. Generally, these models comprise an equivalent feeder, induction motors, electronic and static loads, and one or a few aggregated \ac{DER} models. This approach simplifies the representation of complex \acp{DN} while preserving key electrical characteristics. Two examples are depicted in Fig.~\ref{fig:Dyn_Eq_ADNs}: the \ac{CMPLDWG} proposed by the \ac{WECC} and the \ac{SLMR} proposed by the China Electric Power Research Institute. The detailed governing differential-algebraic equations and control strategies for \ac{CMPLDWG} and \ac{SLMR} can be found in \cite{MaZixiao2020} and ~\cite{TiankaiLan2024}, respectively. Readers are referred to these works for the full mathematical formulations. These models also incorporate partial or total tripping of some components, depending on the magnitude of the voltage. 

The main limitation of these models is that all components are connected to the same bus, which overlooks the voltage diversity in the \ac{DN}. Also, these models fail to capture the interactions and proximity effect of groups of \acp{DER} with nearby motors directly connected to the grid or interfaced with \acp{VFD}. During large voltage dips in the \ac{TN}, some induction motors in the \ac{DN} may stall, which momentarily worsens the local voltage conditions before their thermal tripping is activated. Meanwhile, inverter-based \acp{DER} with voltage-regulation capabilities, such as \ac{DG} units and \ac{BESS}, may need to inject more reactive power if not disconnected due to low-voltage conditions. Similarly, \acp{VFD} could trip due to the low-voltage condition and remain disconnected for several minutes~\cite{SullbergCisco2021}.  Indeed, the disconnection of \acp{DER} and loads during disturbances is a major source of uncertainty, and this should be accounted for in dynamic equivalents. 

Another challenge with the dynamic equivalents of \acp{ADN} within the \textit{gray-box} category~\cite{Milanovic2013} is the model parameterization. Although the components of the equivalent \ac{ADN} are assumed to be known, the exact composition is not. Therefore, parameter identification techniques must be applied using real measurements of past events, which are mostly unavailable, or simulations of a more detailed \ac{DN} model \cite{AlvarezFernandez2020,Chaspierre2021,Rabuzin2022}. 
Alternatively, \textit{black-box} models try to match the input-output relations without any physical knowledge of the \ac{DN}~\cite{Milanovic2013}, for example~\cite{ChaoZheng2019,LDOspina2022}. However, the possibility of inaccurate responses due to disturbances not accounted for during training could diminish engineers' interest. More work is needed to build trust in these models. 

Regardless of whether gray-box or black-box models are used, without real measurements of past events, developing accurate dynamic equivalents of \acp{ADN} will require detailed \ac{DN} models, i.e., agreements on model sharing are necessary; otherwise, the \ac{DSO} must provide the resulting model, in accordance with the model specifications defined by the \ac{TSO}.

\subsection{T\&D Simulation Tools}

Contrary to dynamic equivalents of \acp{ADN}, several works have investigated the use of combined simulations of \ac{TD} networks~\cite{Thakar23}. These simulations provide accurate and high-fidelity results of the system dynamics and the interactions between \ac{TN} and \acp{DN}. 

There are two types of simulation approaches: Combined-type and co-simulations. The combined type integrates the models of all voltage levels within the same simulator~\cite{AristidouPetros2015, ARISTIDOU2015}, assuming balanced operation of the \ac{DN}, suitable for stability studies under the phasor approximation. The \ac{TD} co-simulators process the \ac{TN} model in one simulator and the \ac{DN} model in a second simulator, exchanging information at each integration time step \cite{ZhengLei2024,LiuYuan2024}. Generally, the \ac{TN} simulator provides the voltage at the point of common coupling, and the \ac{DN} simulator provides the active and reactive power demand at the same point \cite{Venkatraman2019}. 

The advantage of \ac{TD} simulations is the possibility of modeling the \acp{DER} and dynamic loads in detail, i.e., not aggregated.  The main limitation is the computational burden, which restricts simulations to specific regions of interest. Still, \ac{TD} simulations are a valuable tool for gaining a better understanding of the dynamic interactions between network levels. They are ideal for assessing the performance of new control schemes of \ac{DER} aggregations or \acp{VPP} and their impact on transmission and distribution grids. Finally, they also serve as a benchmark for assessing the accuracy of dynamic \ac{ADN} equivalents.

\section{TSO-DSO Coordination Mechanisms in Network Planning and Operations}
\label{sec:mechanisms_and_market}
We next detail \ac{TSO}-\ac{DSO} coordination mechanisms along two dimensions: a) coordinated network expansion planning, and b) coordinated operation, covering respectively network expansion planning and look-ahead operational planning. The coordinated market-based procurement of flexibility is discussed later in Section~\ref{sec:market_aspects}. 

We differentiate between two coordination time-scales: 1) the planning timescale, in which flexibility (from different grid levels) is considered in network investment and expansion planning exercises by TSOs and DSOs, and 2) the operational time-scale, which considers TSO-DSO coordination in network operations, including the activation of flexibility to resolve anticipated operational grid issues, such as balancing, congestion management, or voltage control.

\subsection{TSO-DSO Coordination in Network Expansion Planning}
Accounting for generation-side and demand-side flexibility in network expansion planning has been of key recent importance as \acp{SO} aim to manage, optimize, and prioritize their network investments plans, especially in light of the large projected increase in load as well as in local generation. As flexibility is increasingly available from different grid levels, TSO-DSO coordination in network expansion planning becomes ever more important to correctly account and plan in the available flexibility volumes. This topic was studied in FlexPlan~\cite{FlexPlanWebpage}. This was a large-scale European project that aimed to develop a new tool for co-optimizing transmission and distribution grid planning by considering demand-side, generation-side, and storage flexibility within transmission and distribution grids as an alternative to traditional network reinforcements. Six regional cases were studied for different European regions, spanning the three decades of 2030-40-50. 

As peak consumption occurs not only in winter but also in summer due to the widespread air conditioning and \ac{RES}, it is important to analyze a large number of potential operation scenario combinations. Planning analyses cannot readily be carried out using a Monte Carlo approach, i.e., by running many separate simulation scenarios that consider different \ac{RES} generation profiles, as they may provide different and conflicting planning options. For this reason, FlexPlan opted to solve a probabilistic \ac{OPF} to minimize total costs (operational and investment) for a linear combination of all considered scenarios weighted with their relevant probabilities. However, solving such a problem for an exceedingly large system encompassing thousands of nodes, including both \ac{TN} and \acp{DN}, and hourly dispatch resolution is not tractable, prompting the project to apply Benders’ decomposition\cite{ShahidehpourFu2005}, implementing a natural model decomposition pertaining to \ac{TN} and \acp{DN}. This decomposition is motivated by the different physical characteristics (R/X ratios) and topologies (meshed in \ac{TN}, whereas mainly radial in \ac{DN}), which require different modeling hypotheses. Furthermore, the possibility that transmission and distribution grids can provide mutual services motivates a hypothetical integrated planning, in which \ac{TSO} and \acp{DSO} together elaborate a common set of scenarios and agree on a list of proposed interventions. Such an integrated approach has a theoretical optimality advantage. However, beyond the large dimension of the resulting model, another challenge consists in the need to  integrate and coordinate a large number of \acp{DSO} (e.g., in Italy alone, there are more than 300 \acp{DSO} with varying levels of availability of tools, methodologies, etc.). This scalability challenge of the common planning approach is added to the underlying challenges related to sharing internal data of \acp{SO}, which raises privacy issues. 

A practical approach consists of implementing a coordinated planning methodology, in which \ac{TSO} and \acp{DSO} exchange only data on the border between their jurisdictions. Yet, each operator pursues an autonomous planning calculation, in which the neighboring grids are modeled via simplified equivalents. 
The FlexPlan project developed a methodology implementing such a coordinated approach \cite{MRossini2023}. However, a still unresolved problem lies in finding a linear yet accurate (equivalent) model of a \ac{DN}, an important precondition for ensuring numerical tractability by keeping these large optimization models linear~\cite{MigliavaccaEtAl2022}. As the definition of the equivalent model changes with the operating point for which it is calculated, the choice of the best compromise remains a delicate aspect of this methodology. Planning decisions also span multiple time scales: long-lead investments can often be assessed with linearized grid models, whereas short-horizon deployments of easily installed assets may require more detailed power-flow representations to capture voltage and reactive power effects. TSO–DSO coordinated expansion planning, and grid modeling practices therein, should therefore explicitly accommodate these differing time horizons and their modeling granularity requirements.

\subsection{TSO-DSO Coordination for Power System Operation}
Many mechanisms to coordinate \ac{TSO} and \ac{DSO} operations have been proposed \cite{ZhaoYuan2017,AMohammadi2019,KunjieTang2021,Bakhtiari2023,MikhailBragin2022,LuKaicheng2024,WengYu2023,TianceZhang2023,MUsman2023} since the topic inception almost a decade ago. These mechanisms can be classified into two groups \cite{MUsman2023}:
\begin{itemize}
\item Group A: Mechanisms that look at \ac{TSO} and \acp{DSO} control areas as an entire system and minimize its overall operation cost i.e., summing up the cost of \ac{TSO} and \acp{DSO} \cite{ZhaoYuan2017,AMohammadi2019,KunjieTang2021,Bakhtiari2023,MikhailBragin2022}. 
\item Group B: Mechanisms that recognize the peculiarities and data privacy of \acp{TSO} and \acp{DSO} and treat them as autonomous and competing entities in optimizing their objectives \cite{LuKaicheng2024,WengYu2023,TianceZhang2023,MUsman2023}. For example, typically, a regulatory body approves investment plans for these two monopolistic entities. Operating expenses are generally covered by grid-access tariffs and a limited, constrained common fund. This creates competition between the regulated monopolies, which may reduce their willingness to coordinate and maximize social welfare.
\end{itemize}

The works of group A can be applied in some contexts (e.g. vertically integrated systems) but are impractical in deregulated market environments (e.g. Europe, North-America) since: (i)  \acp{TSO} and \acp{DSO} have their own specific goals, limited to their control area, and compete for utilizing \ac{DER}s’ flexibility to minimize their cost, (ii) to preserve data privacy and sensitive information, \acp{TSO} and \acp{DSO} cannot always share their cost or grid models, and (iii)  \acp{TSO} and \acp{DSO} have a mandate and a duty to operate their grids safely, which they cannot readily pass along to a coordinating third party agent. 

Most works of group B, e.g., \cite{LuKaicheng2024,WengYu2023} are computationally heavy as they need many (non-parallelizable) iterations between the separate solutions of cost-optimization problems for \acp{TSO} and \acp{DSO}, while the time for solving a problem at either the \ac{TSO} and \ac{DSO} sides is not insignificant. Furthermore, reaching a consensus (e.g., a Nash equilibrium in case of competitive interdependent environments) between \ac{TSO} and \acp{DSO}, e.g.,  \cite{LuKaicheng2024} is not always guaranteed.
Two distinct methodologies proposed in group B \cite{TianceZhang2023,MUsman2023}  are non-iterative, i.e. exhibit only a few steps. This non-iterative nature presents some practical advantages (e.g., efficient coordination, as only a small number of \ac{TSO} and \ac{DSO} problems are solved, less interactive information, non-oscillatory convergence issues). 

Regarding solution approaches, most methodologies proposed for TSO-DSO coordination employ distinct decomposition methods (e.g. hierarchical \cite{ZhaoYuan2017}, Benders’ decomposition \cite{Bakhtiari2023}, surrogate Lagrange relaxation \cite{MikhailBragin2022}, alternating direction method of multipliers \cite{MarquesLuciana2023}). However, such approaches usually necessitate many iterations between the \ac{TSO} problem and \acp{DSO} problems, being computationally burdensome. Some works rely on convex relaxations (e.g. second order \cite{Papavasiliou2018,ZhaoYuan2017,MikhailBragin2022}). However, these relaxations do not compute feasible solutions of the original problem, except for very specific conditions, which are difficult to meet for more complex practical grid models. For example, in some works, the \ac{DSO} problem is even oversimplified using a linear (DC) power flow approximation (typically employed at transmission level), which poses accuracy challenges. 
Table~\ref{tab:attributes_tso_dso_coord} synthesizes the key attributes of these works in terms of modelling complexity, scalability and the size of transmission and distribution systems used for testing. 

\begin{table*}[ht]
\centering
\renewcommand{\arraystretch}{1.2}
\caption{The main attributes of existing TSO-DSO coordination methods}
\label{tab:attributes_tso_dso_coord}
\begin{tabularx}{\textwidth}{|c|Y|Y|Y|Y|Y|Y|Y|Y|Y|c|c|}
\hline
\multirow{2}{*}{\textbf{Ref.}} &
\multicolumn{2}{c|}{\textbf{Type of flexibility}} &
\multicolumn{2}{c|}{\textbf{Optimization model}} &
\multicolumn{3}{c|}{\textbf{Model attributes}} &
\multicolumn{2}{c|}{\textbf{Cost of flexibility}} &
\multirow{2}{*}{\makecell{\textbf{Grid size}\\\textbf{TSO/DSO}}} &
\multirow{2}{*}{\textbf{Scalable}} \\
\cline{2-10}
& \textbf{Active power} & \textbf{Reactive power}
& \textbf{TSO} & \textbf{DSO}
& \textbf{Stocha-sticity} & \textbf{Multi-period} & \textbf{N-1 cont.}
& \textbf{Active power} & \textbf{Reactive power}
& & \\
\hline
\cite{ZhaoYuan2017}  & * & * & SOCP & SOCP &   &   &   &   &   & 6 / 336   & * \\
\hline
\cite{AMohammadi2019}  &   &   & DC   & NLP  &   &   &   &   &   & 118 / 2   & * \\
\hline
\cite{KunjieTang2021}  &   &   & NLP  & NLP  &   &   &   &   &   & 118 / 69  &   \\
\hline
\cite{Bakhtiari2023}  & * &   & DC   & LP   &   &   &   &   &   & 118 / 33  & * \\
\hline
\cite{MikhailBragin2022}  & * & * & LP   & SOCP &   &   &   &   &   & 118 / 34  & * \\
\hline
\cite{LuKaicheng2024}  & * & * & NLP  & NLP  &   &   &   &   &   & 30 / 69   &   \\
\hline
\cite{WengYu2023} & * &   & NLP  & LP   &   & * & * &   &   & 9 / 141   &   \\
\hline
\cite{TianceZhang2023} & * &   & LP   & LP   &   & * & * &   &   & 118 / 141 &   \\
\hline
\cite{MUsman2023} & * & * & NLP  & NLP  & * & * & * & * & * & 60 / 34   &   \\
\hline
\end{tabularx}
\end{table*}

Distributed optimization naturally aligns with the decoupled physical structure of TSO-DSO systems. However, its practical implementation faces significant hurdles. Distributed leader-follower frameworks have demonstrated that distributed algorithms can effectively manage \ac{OPF} and stability assessment \cite{LiZhengshuo2019}. These methods offer proven optimality and competitive convergence rates. Nevertheless, serious barriers persist in multi-period settings. The time-coupling of energy storage and flexible demand complicates the derivation of aggregated flexibility cost curves~\cite{FlCapitanescu2025}. Furthermore, the inherent non-linearity and non-convexity of network models often lead to slow convergence and excessive iterative overhead. For time-critical operations, centralized or semi-centralized approaches remain more robust. Future coordination must therefore strike a balance between local autonomy and the necessary centralized oversight to ensure system-wide security. Still, hierarchical controls in time and space (i.e. the three-layer model—tertiary, secondary, and primary) must be adapted to the new coordination mechanisms. 

As surveyed in Table~\ref{tab:attributes_tso_dso_coord}, three major research gaps requiring meaningful research can be noticed: 
\begin{itemize}
\item In terms of models, most works address over-simplified and inaccurate problems. The work in \cite{MUsman2023} has tackled an exceedingly challenging \ac{TSO}-\ac{DSO} coordination model. Indeed, the methodology includes essential yet challenging practical operation requirements, i.e. the N-1 security criterion at \ac{TSO} level, operation uncertainties and time coupling at both \ac{TSO} and \ac{DSO} levels, and emerging flexible assets (energy storage devices and flexible demand). This leads to very large \ac{NLP} problems to be solved, posing computational challenges to be resolved. 
\item There is no approach that retains tractability in the presence of the most computationally heavy model features (stochasticity, multi-periods, N-1 security). The sizes of transmission and distribution systems also play a key scalability role, as most works in the literature employ test systems that are  much smaller than in practice. Note that, some approaches in Table~\ref{tab:attributes_tso_dso_coord} are marked as not scalable as, despite relying on scalable problems (e.g. \ac{LP}), the number of problems to be solved in series, corresponding to iterations between \ac{TSO} and \acp{DSO}, is large in practice (e.g. tens to few hundreds); 
\item Very few works consider the cost of aggregated flexibility of an \ac{ADN} at the \ac{TSO}-\ac{DSO} interface. 
\end{itemize}

In summary, despite extensive research on \ac{TSO}-\ac{DSO} coordination, the lack of both accuracy and tractability of state-of-the-art methodologies prevents their practical application.  Furthermore, automatic closed-loop control schemes, such as \ac{OFO}, can help manage system complexity by embedding real-time measurements into the optimization loop~\cite{ZettlIrina2026}. 

\section{TSO-DSO Coordination for the Market-based Procurement of Flexibility}
\label{sec:market_aspects}

\subsection{TSO-DSO Coordinated Flexibility Market Models}
As the need for flexibility is increasing at both \ac{DSO} and \ac{TSO} levels, and the sources of flexibility are growing in availability at both transmission and distribution levels, there is a key need for \ac{TSO}-\ac{DSO} coordination, in particular, for the pre-qualification, procurement, and activation of flexibility to (i) ensure the efficiency of the procurement process, maximizing the value-stacking potential of flexibility, whereby a flexibility volume can simultaneously serve the needs of multiple \acp{SO}, and (ii) ensuring that any flexibility activated by one \ac{SO} from a flexibility source can be done in a grid-safe\footnote{Grid-safety here refers to the activated flexibility not leading to other congestions or voltage issues in any of the impacted grids.} way, even in the cases where the \ac{SO} activates flexibility from sources connected outside its own area of control. The coordination of the flexibility procurement process via flexibility market mechanisms has led to several \ac{TSO}-\ac{DSO} coordination schemes that were first proposed as part of the SmartNet project \cite{SmartNetWebpage}, and then further evolved and refined as part of the follow-up CoordiNet \cite{CoordiNetWebpage} and finally OneNet \cite{OneNetWebpage} projects\footnote{The SmartNet, CoordiNet, and OneNet projects are all large-scale European projects that have tested, via pilot activities, the feasibility and efficiency of various TSO-DSO coordinated flexibility market models.}. Hereafter, we refer to these schemes as \ac{TSO}-\ac{DSO} coordinated flexibility market models\cite{MarquesLuciana2023}. 

These schemes start from \textbf{disjoint markets}\cite{Sanjab2021_SOCP,Papavasiliou2018,SanjabAnibal2022,MarquesLuciana2023} setting, namely a transmission-level disjoint market and a distribution-level disjoint market, in which each \ac{SO} procures flexibility to resolve its own needs and without any access to flexibility from grids outside their area of control. That covers not only the direct access to flexibility bids, but also indirect access through caused imbalances and changes to interface flows. In other words, under the disjoint market schemes, the flexibility procurement performed by each \ac{SO} is done in a balanced way so that it does not impact the operation of other grids. As there is no exchange of flexibility (i.e., each SO procuring flexibility solely from resources connected to its own grid), each \ac{SO} has to account only for its own grid constraints when clearing those markets. 

On the opposite side of the spectrum, the \textbf{common market} model\cite{SanjabAnibal2022,MarquesLuciana2023} is proposed, wherein \acp{TSO} and \acp{DSO} jointly procure flexibility (in a co-optimized manner) to collectively resolve their needs from a common pool of flexibility resources (originating from the different grids) while abiding by the constraints of all the grids involved. This leads to a single joint optimization problem to be solved to meet the needs for all SOs at the least possible costs. Two other proposed \ac{TSO}-\ac{DSO} coordination schemes fall within this spectrum, and are two sequential market schemes, namely, the \textbf{multilevel market} and the \textbf{fragmented market} models. 

The multilevel market\cite{MarquesLuciana2023} is a sequential, two-stage structure. In Layer 1, each \ac{DSO} procures flexibility from \acp{FSP}  connected to its distribution grid, subject to local operational constraints. In Layer 2, the \ac{TSO} procures flexibility to address transmission-level needs, using both transmission-connected resources and the unused portions of bids forwarded from Layer 1\footnote{We note here that distribution-levels bids are not aggregated and can therefore be individually cleared by the \ac{TSO}. In addition, for bid forwarding to be feasible, the product characteristics in the two market layers should be compatible.}.  This design grants \acp{DSO} priority access to distribution-level flexibility, while allowing the \ac{TSO} to use any remaining capacity. For this flexibility to be used in a grid-safe manner for the distribution grids, the \ac{TSO} must then account for the \ac{DN} constraints when clearing its market. 

The model features two forms of flexibility sharing:
\begin{itemize}
\item	Direct sharing: a \ac{SO} procures flexibility from resources located outside its grid. In particular, in Layer 2, the \ac{TSO} directly accesses distribution-connected flexibility bids forwarded from \acp{DSO}.
\item	Indirect sharing: \ac{SO} actions in its own grid create system needs resolved in a subsequent market. In Layer 1, \acp{DSO} do not procure transmission-connected bids. However, their flexibility procurement can modify the net power exchange at the \ac{TSO}-\ac{DSO} interface, creating imbalances for the \ac{TSO}. These are subsequently addressed in Layer 2 using available transmission- and distribution-level bids. Thus, Layer 1 enables indirect flexibility sharing between \acp{DSO} and the \ac{TSO}.
\end{itemize}

The \textbf{fragmented market}\cite{MarquesLuciana2023} is also a sequential two-layer design. Layer 1 mirrors the first layer of the multilevel market: each \ac{DSO} operates an independent market to procure flexibility from resources connected to its own grid, subject to local constraints. This procurement may modify the \ac{DSO}’s net interchange with the \ac{TSO}, providing \acp{DSO} with indirect access to transmission-level flexibility, as resulting imbalances are resolved in Layer 2 (or in European balancing platforms), unless they are naturally netted out.
Layer 2, however, differs fundamentally from the multilevel design. In the fragmented market, the \ac{TSO} does not receive unused distribution-level bids from Layer 1 and therefore cannot access distribution-connected flexibility. Consequently, while DSOs retain indirect access to \ac{TSO}-level flexibility through imbalance adjustments, neither the \ac{TSO} nor the \acp{DSO} has direct access to flexibility bids from other grid levels. As such, an \ac{SO} must only account for its own network constraints when clearing their market layers.

A summary of the features of these different \ac{TSO}-\ac{DSO} coordinated market models is shown in Table~\ref{tab:coord_flex_mark}.

The market clearing problem in each of these markets practically boils down to an optimization problem in which a \ac{MO} aims to choose the set of bids, or portions thereof, to clear (quantities being decision variables) to minimize the total procurement costs (cost terms are submitted bid prices) while abiding by all imposed network constraints (i.e., resolving original congestion, balancing needs, voltage issues while not creating any other constraint violations) as well as the bid constraints. Indeed, to represent \acp{FSP}’ technical and economic constraints when providing flexibility, different bid\footnote{In Flexibility markets bids can be submitted by aggregators of different \acp{DER} or by individual users depending on the market rules and participation requirements (minimum bid size, full activation time constraints, among others).} formats of increasing complexity are typically considered in flexibility markets. These included, e.g.,: fully-divisible bids that can be cleared anywhere from 0 up to the offered maximum quantity (thus, not including any minimum bid clearing quantity requirement); indivisible bids, that can only be cleared at the offered volume (all-or-nothing); partially divisible bids (clearable between a minimum quantity and the offered maximum), in addition to other more complex bid formats which impose logical conditions across multiple bids. Complex formats include exclusive bids, where at most one bid in a set can be accepted, and multipart (parent–child) bids, where a child bid can be cleared only if its parent is at least partially cleared. While these formats improve the expressiveness of flexibility offers, their logical constraints typically require integer and binary decision variables in market-clearing models, increasing computational complexity.

The market clearing formulations for the different TSO-DSO coordinated markets differs based on the bids available in each market layer, the constraints on the interface flow, and the different network constraints to be considered. 

It is important to note that these \ac{TSO}-\ac{DSO} coordinated market models aim to coordinate the procurement of flexibility between \acp{TSO} and \acp{DSO} to maximize procurement efficiency and minimize risks to the impacted grids. These models can be applied to different forms of flexibility products, which can differ in their time scale from long-term to short-term and near real-time. An overview of different flexibility market products is presented in \cite{OneNetD22} and their application in a regional flexibility market demonstration activity across Finland, Estonia, Lithuania, and Latvia as part of the Horizon Europe OneNet project~\cite{OneNetWebpage} is presented in~\cite{OneNetD74}. Nonetheless, in such market-based flexibility mechanisms, flexibility is procured before real time (shortly before, in case of near-real-time market segments), to be activated during real-time operation. As such, the element of uncertainty on the availability of flexibility, from the \ac{FSP}'s side, and uncertainty regarding the network state evolution and hence the flexibility need, from the \ac{SO}'s side, are important in defining how an \ac{FSP} bids in the market and how \acp{SO} define their flexibility needs to be procured via the market. In addition, stochastic forms of TSO-DSO coordinated market clearing formulations can be also proposed in this respect, akin to uncertainty-based market clearing formulation for wholesale markets.

Next, we compare between these different TSO-DSO coordinated flexibility market models by focusing on: 1) How do their efficiencies compare and the drivers thereof (Sec.~\ref{subsubsec:MarketModelComparison}), 2) stability of cooperation between the \ac{TSO} and \acp{DSO} in a joint market and cost allocation schemes for the jointly procured flexibility (Sec.~\ref{subsubsec:StableTSODSOCooperation}), 3) how network constraints can be accounted for in the market clearing models (Sec.~\ref{subsubsec:MarketNetworkConstraints}), 4) risks of FSP strategic bidding in each and impacts thereof (Sec.~\ref{subsubsec:StrategicBehavior}). 

\begin{table}[!htbp]
\centering
\caption{TSO-DSO Coordinated Flexibility Markets}
\label{tab:coord_flex_mark}
\resizebox{\linewidth}{!}{%
\begin{tabular}{|
>{\centering\arraybackslash}m{0.24\linewidth}|
>{\centering\arraybackslash}m{0.13\linewidth}|
>{\centering\arraybackslash}m{0.13\linewidth}|
>{\centering\arraybackslash}m{0.16\linewidth}|
>{\centering\arraybackslash}m{0.18\linewidth}|
>{\centering\arraybackslash}m{0.16\linewidth}|}
\hline
\multicolumn{3}{|c|}{\textbf{Market}} &
\multicolumn{3}{c|}{\textbf{Flexibility \& Network Information Sharing}} \\
\hline
\textbf{Model} & \textbf{Clearing} & \textbf{Stages} &
\textbf{TSO resource} & \textbf{DSO resource} & \textbf{Network info.} \\
\hline

Disjoint TN-level
& \multirow{2}{*}{Indep.}
& \multirow{2}{*}{N/A}
& \multicolumn{3}{|c|}{\multirow{2}{*}{No}} \\
\cline{1-1}
Disjoint DN-level
&  &  & \multicolumn{3}{|c|}{} \\
\hline

Common
& Joint
& 1
& \multicolumn{2}{|c|}{Complete sharing}
& Yes \\
\hline

Fragmented
& \multirow{2}{*}{Sequential}
& \multirow{2}{*}{2}
& \multirow{2}{*}{Indirect}
& No
& No \\
\cline{1-1}\cline{5-6}
Multilevel
&  &  &
& Direct (w.\ DSO priority)
& Yes~(Layer~2) \\
\hline
\end{tabular}%
}
\end{table}

\subsection{Market Model Efficiency  Comparison}\label{subsubsec:MarketModelComparison}
Several dimensions can be considered when comparing between the presented market models. Nonetheless, one of the key comparison components is their efficiency, i.e., the cumulative costs of procuring flexibility to resolve the different grid needs, as this has a direct impact on consumers' bills and on the ability of \acp{SO} to economically and practically rely on flexibility mechanisms to support (and at instances, delay or even replace) network investments.  
The economic efficiency of the aforementioned \ac{TSO}-\ac{DSO} coordinated market models, focusing on their ability to procure flexibility at minimum system cost, was extensively analyzed in \cite{MarquesLuciana2023}. The common market achieves the theoretically highest efficiency, as all \acp{SO} jointly procure flexibility from a unified resource pool under full network constraints. However, this market structure faces network sharing information and governance challenges that hinder its practical implementation. In terms of efficiency, several structural factors influence relative efficiencies across the models.

A central driver is the pricing of the \ac{TSO}-\ac{DSO} interface flow \cite{MarquesLuciana2023,Sanjab2023_SequentialTSODSO,FlCapitanescu2025,FlorinCapitanescu2024}, which governs the cost of indirect flexibility sharing in sequential designs. As shown in~\cite{MarquesLuciana2023} When the interface flow is unpriced, multilevel and fragmented markets tend to procure unnecessary downward flexibility in their local layers, creating costly imbalances subsequently corrected by the \ac{TSO}. Appropriate interface pricing mitigates this inefficiency. Indeed, as proven in~\cite{MarquesLuciana2023}, under optimal pricing (a pricing reflecting the dual variables of the interface flow constraint), both multilevel and fragmented markets can theoretically match the efficiency of the common market, although deriving such optimal prices may be challenging in practice (this element is further elaborated in Sec.~\ref{subsubsec:MarketNetworkConstraints}.

Entry barriers, such as minimum bid sizes, also significantly affect efficiency. While the common market benefits from full pooling and co-optimization, stringent participation requirements may exclude small-scale resources~\cite{Sanjab2023_SequentialTSODSO}. In contrast, multilevel and fragmented schemes enable local participation in Layer 1, partly offsetting the efficiency loss caused by market fragmentation. However, in fragmented and disjoint distribution-level markets, locally participating resources cannot serve \ac{TSO} needs, limiting overall efficiency gains. As a result, efficiency rankings depend on the specific level of the resulting entry barriers. 

The ability for an \ac{FSP} to aggregate flexibility resources is impacted by the market design and TSO-DSO coordination scheme in place. Aggregation enables an \ac{FSP} to combine multiple flexibility resources into a portfolio capable of delivering system and grid-level services. This facilitates compliance with market requirements (e.g., minimum bid sizes, response times) while allowing the \acp{FSP} to further improve the economic optimization of flexibility offering and delivery, thereby enhancing their market participation potential, as well as the liquidity and efficiency of the flexibility market, as a result. The separation between market layers –- i.e., the transition from common to multilevel, fragmented, or disjoint market structures –- can affect the aggregation potential of \acp{FSP}. However, aggregation feasibility depends not only on the market design/TSO-DSO coordination scheme, but also on grid operational conditions and the locational requirements of the procured flexibility service. For system-wide services such as balancing, aggregation across resources connected at different distribution levels can be feasible when local network constraints are non-binding, as such services require limited locational granularity. Conversely, when local constraints become active, detailed locational information is required to ensure that flexibility activation does not compromise distribution-grid security. This is reflected in the common market and in Layer 2 of the multilevel market, where distribution-grid constraints must still be represented in the market-clearing process even when the procured service is intended for the \ac{TSO}. Similar requirements arise for grid services such as congestion management and voltage control, which inherently rely on detailed network-location information. Consequently, although common markets are structurally more supportive of aggregation, active local constraints and location-sensitive services can significantly limit aggregation possibilities, even under fully integrated market designs.

Bid formats further influence efficiency outcomes as investigated in~\cite{Sanjab2023_SequentialTSODSO}. The common market is relatively robust to minimum clearing requirements introduced by partially divisible bids due to its larger resource pool. Fragmented and sequential models, with smaller and tiered pools of flexibility bids, exhibit greater efficiency losses when such constraints restrict substitution options, e.g., a cheap bid with high minimum clearing requirements can be cleared in a common market due to the large common flexibility needs, but might not be cleared in any of the sequential layers of the multilevel market.

\subsection{Stable TSO-DSO Cooperation and Cost Allocation}\label{subsubsec:StableTSODSOCooperation}

As discussed, collectively, the common market can return a minimum-cost solution to all participating \acp{SO}. However, here, several questions arise: (i) Would each participating \ac{SO} be individually better off joining this grand coalition  (cooperation between all \acp{SO}) as compared to acting alone (disjoint market) or forming a sub-coalition? (ii) When joining this grand coalition and jointly procuring flexibility, how should the costs of this collectively procured flexibility split among the participating \acp{SO}? (iii) What are the potential cost allocation methods that can be implemented and what are their mathematical properties and practical implications? These challenges were addressed in \cite{SanjabAnibal2022}, which casts the \ac{TSO}–\ac{DSO} cooperation as a cooperative game, namely, as a characteristic-function cost game, where the value of each coalition corresponds to the minimum flexibility procurement cost under the associated sub-common or common market.

The analysis~\cite{SanjabAnibal2022} proved that the game takes the form of a linear production game, enabling the application of classical cooperative-game results. A central theoretical contribution is the proof of non-emptiness of the Core, implying that the grand coalition of all SOs is stable, and that cooperation naturally arises without external incentives. Multiple cost-allocation mechanisms were investigated, and \ac{SV} emerged as the theoretically strongest mechanism. The \ac{SV} for joint TSO-DSO cost allocation has also been explored in \cite{TSODSOCooperation2019}, which analyses coordination mechanisms among \acp{TSO}, \acp{DSO}, and retailers for the procurement of distributed flexibility. 
However, the \ac{SV} suffers from high computational complexity (being NP-complete), which can limit its practical implementation at instances. Other methods \cite{SanjabAnibal2022}, such as the Lagrangian-based allocation, Equal-Profit method, and proportional cost allocation, could provide adequate practical alternatives when \ac{SV} implementation is prohibitive in terms of computational cost, albeit they miss a few mathematical properties (e.g., additivity and anonymity properties). A key finding of \cite{SanjabAnibal2022} is highlighting that different cost-allocation mechanisms can generate significant disparities in how costs are distributed among \acp{SO}, with some methods shifting the burden more heavily onto specific parties (\acp{DSO} or \acp{TSO}) despite identical market conditions, necessitating proper negotiation frameworks for the setup of common markets.

\subsection{Network Constraints Considerations}\label{subsubsec:MarketNetworkConstraints}

When considering the use of distributed flexibility in \ac{TSO}-level markets, namely, in Layer 2 of a multilevel market, or in centralized schemes similar to the common market, the challenge always arises with respect to ensuring that the procurement and activation of distributed flexibility in centralized markets do not lead to \emph{constraint violations} in \acp{DN}, where this flexibility is located. Beyond the overarching key challenge of including complex distribution-level network representations in the market clearing constraints (which is addressed in Section~\ref{subsubsec:MarketChallenges}), this presents a heightened challenge, as \acp{TSO} typically lack practical access to \ac{DN} models and cannot practically include distribution-system constraints in their market-clearing formulations.  Several approaches have been proposed in the literature to address this challenge. We next explore three different solutions thereof: (1) distributed market-clearing approaches, (2) operating envelopes as a proxy to full network representation, and (3) prequalification and sequential market structuring.   

\subsubsection{Distributed/Decentralized Solutions}

\begin{figure}[h!]
    \centering
    \includegraphics[width=0.4\textwidth
    ]{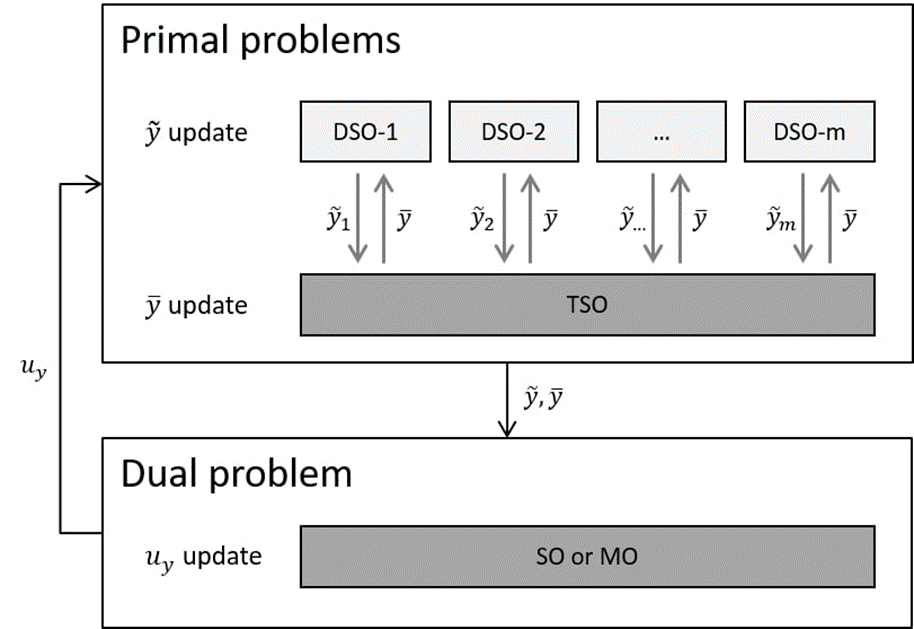}
    \caption{ADMM mechanism for distributed market clearing (figure replicated from \cite{MarquesLuciana2023})}
    \label{fig:ADMM_mech}
\end{figure}

To avoid the need for network information sharing, the work in \cite{MarquesLuciana2023} has proposed the use of distributed solutions to the common and multilevel markets, in which each \ac{SO}/\ac{MO} solves a local problem (similar to the disjoint markets, thus not requiring network information from external grids) and an iterative mechanism is put in place based on the \ac{ADMM} to converge on the optimal interface quantities and prices. This \ac{ADMM} mechanism is highlighted in Fig.~\ref{fig:ADMM_mech}. 

This approach also allows the calculation of \emph{optimal interface flow pricing} in a distributed manner, which is a feature of prime importance, as when the interface flow is priced optimally, the solutions of the common market, multilevel market, and fragmented markets are proven to converge (thus not requiring the sharing of network information) as shown in \cite{MarquesLuciana2023} and discussed in Section~\ref{subsubsec:MarketModelComparison}. This method, however, requires \ac{ICT} in place for \acp{TSO} and \acp{DSO}, allowing them to exchange interface flow quantity and prices calculations to converge to the optimal and grid-safe solution. In addition, even when theoretical requirements for convergence are met\cite{boyd2010distributed}, \ac{ADMM} can require many serial iterations between the \ac{TSO} and the \ac{DSO}s’ problem, which may become time-prohibitive for practical deployment. When these \ac{ICT} capabilities prove to be challenging, other non-iterative approaches, as introduced next, can provide practical alternatives.  

\subsubsection{Flexibility Region and Operating Envelopes Integration in Market Clearing}
\label{sec:FR_integration_market}

A proxy for incorporating full \ac{DN} models/constraints at the transmission-level market can be achieved using flexibility regions at the \ac{TSO}-\ac{DSO} interface (introduced in Section~\ref{sec:feasible_regions}) -- which can include static limits or dynamic $(P, Q, V)$ flexibility curves to provide active voltage regulation and \acp{AS} in a \ac{VPP}-like manner~\cite{GridCognizantTSODSO2024} --, or through P/Q limits at each user connection points, which typically fall within the realm of what is defined as (dynamic) \acp{OE}~\cite{petrou2021ensuring,liu2021grid,liu2022using,Riaz2019}. 
Generally, \acp{FR} and \acp{OE} specify safe power ranges at the different grid connection points, thereby aiming to ensure that when inside the specified PQ chart, the flexibility actions remain grid-safe for the \ac{DN}. From a flexibility markets perspective, these \acp{FR}/\acp{OE} can then be used in \ac{TSO}-level flexibility market formulations \cite{MUsman2023,Marques2024_OE,Kaushal2024_OE}. In other words, \ac{FR}/\acp{OE} can be computed at the pre-qualification stage (i.e., prior to the market runs), generating P/Q power ranges which can then be included in the flexibility market clearing problem, e.g., in Layer 2 of the multilevel market or even in the common market to avoid the need for the \ac{DSO} to share network information with the \ac{TSO} or with third-party \acp{MO} (wherein, only the permissible power ranges would be shared), while still aiming at ensuring the distribution grid safety of the flexibility activation. Indeed, the integration of \acp{OE} in transmission-level markets has been particularly analyzed in \cite{Marques2024_OE,Kaushal2024_OE}, in which several \ac{OE} calculation methods have been explored and compared in terms of their grid-safety guarantees and impact on market efficiency, showcasing a tradeoff between these two dimensions. Indeed, the tighter the imposed limits, the lower the risks of grid-constraint violations. However, this would lead to conservative solutions that discard an unnecessarily high volume of flexibility, resulting in increased procurement costs, as well as fairness and transparency challenges for participating \acp{FSP}. The works in \cite{Marques2024_OE,Kaushal2024_OE}
also showcase that not all \ac{OE} methods can guarantee grid safety. Indeed, their performance depends on the calculation method used but also the power flow model employed in the calculation. For ensuring grid-safety, an increasingly complex grid model might need to be used (from linear to convex to nonlinear AC formulations, up to unbalanced 3-phase formulations to really capture the distribution grid power flows), which hinders the scalability of the \ac{OE} computation.

\textit{\textbf{c)	Prequalification Measures and Market Design Reorganization}}

As the computation of \acp{FR}/\acp{OE} can be challenging in practice, alternative methods for enabling the \ac{DN} safe use of distributed flexibility in transmission-level markets have been proposed in large-scale demonstration projects and academic literature, which have been modeled and analyzed in \cite{Ananduta2025_BidForwarding}. 
   
\begin{figure}[h!]
    \centering
    \includegraphics[width=0.48\textwidth
    ]{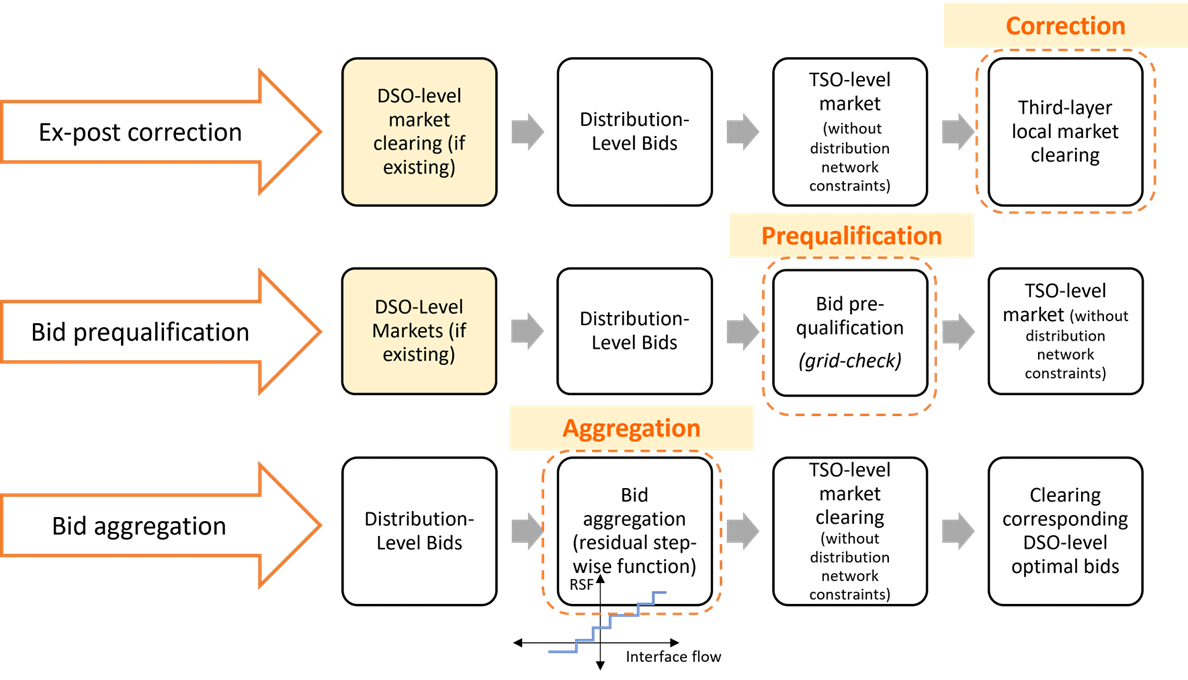}
    \caption{Methods for safe use of distributed flexibility in transmission-markets}
    \label{fig:pre-qual}
\end{figure}

Three solution methods are examined in \cite{Ananduta2025_BidForwarding}, namely: 
(i) a three-layer ex-post corrective market scheme, ii) a bid pre-qualification method, and (iii) a bid aggregation method. These three methods are depicted in Fig.~\ref{fig:pre-qual}. 

The first method, a three-layer corrective market scheme, mitigates the impact of unsafe bids on \acp{DN} by adding a corrective local-market layer after transmission-level clearing. In this design, all remaining bids from the first layer are forwarded to the second-layer \ac{TSO} market, whose market-clearing formulation does not consider \ac{DN} constraints, which may potentially lead to constraint violations. To correct for these violations, once both layers have been cleared, \acp{DSO} run an additional local market to procure any extra flexibility needed to resolve distribution-level congestion induced by the second-layer flexibility procurement. This approach essentially acts as a redispatch stage and mirrors the structure of the Layer-1 market, but with updated cleared quantities and a fixed interface flow taken from Layer 2.

The second method is a bid pre-qualification\footnote{Pre-qualification is a typical step in flexibility markets in which the \ac{SO} checks whether the \ac{FSP} can effectively deliver the service and abide by the product requirements of the service procured. Pre-qualification tests can focus on elements such as \ac{ICT} requirements, technical capability to deliver the service requirements, impacts on the grid, among others.} approach, in which \acp{DSO} filter the bids that can be forwarded to the \ac{TSO} market. The objective is to identify, ex-ante, the subset of bids that can be cleared at the transmission level without creating local congestion. As such, only pre-qualified bids (or portions thereof) are then forwarded to the transmission-level market, which does not then include \ac{DN} constraints in its market-clearing formulation. 

In the third method, bid aggregation, rather than forwarding individual bids, each \ac{DSO} sends step-wise aggregates of feasible flexibility, expressed as changes in the interface flow, to the transmission-level market (Layer 2)\footnote{This scheme is labeled local market model in SmartNet.}, which is relevant in scope to  \cite{FlorinCapitanescu2024,FlorCapitanescu2025}. Each step in this step-wise function, dubbed the \ac{RSF}, represents a grid-safe level of flexibility that the \ac{DSO} can deliver, along with an associated price derived from an optimization problem that considers its own grid constraints. \acp{RSF} are generated by solving the first-layer market problem repeatedly for different fixed interface flow values.  Earlier formulations priced these steps using dual values of the interface-flow constraint, making them less-transparent and potentially sub-optimal, whereas the formulation in \cite{Ananduta2025_BidForwarding} has also adopted price steps using the \ac{DSO}’s optimal primal costs, offering a more transparent interpretation.

The three approaches differ substantially in performance and regulatory compatibility. The three-layer corrective approach is computationally light and aligns with existing regulatory practices, but only guarantees grid safety when sufficient distribution-level flexibility is available. The pre-qualification method reliably filters unsafe bids but may yield suboptimal outcomes. The bid aggregation approach generally provides grid-safe market outcomes with the highest efficiency. However, it is computationally more demanding and potentially challenging from a regulatory standpoint, as it requires \acp{DSO} (or designated third parties) to aggregate and forward bids on behalf of the \acp{FSP}, which faces regulatory challenges. 

\subsection{\ac{FSP} Strategic Bidding}\label{subsubsec:StrategicBehavior} 

The efficiency analysis and comparison between the different \ac{TSO}-\ac{DSO} coordinated market models has focused on the market formulation itself. Nonetheless, this efficiency is also directly impacted by how \acp{FSP} may strategically bid in different \ac{TSO}-\ac{DSO} coordinated flexibility markets. This bidding behavior can change based on the market scheme in place, thus yielding different impacts for different market models. Indeed, an efficiency analysis must also explore the possible exercise of market power~\cite{geckil2016applied}, which can be a key factor of extra costs (and inefficiency) of a market. In the traditional wholesale market setting, market power is a strategic action carried out by single market participants or by a colluding set of participants that aim to reduce the amount of generation offered to the market, or to offer it at a price above the marginal cost of generation, with the aim of influencing market prices and drawing extra profits. In a sequential market setting, an additional opportunity of market power resides in players strategically bidding in the first market layer to create an artificial need that they themselves can resolve in the subsequent market, thus reaping financial benefits (this setting is known as \textit{inc-dec gaming}) \cite{IncDecOriginal}. 
Indeed, several works in the literature have explored market player strategic bidding in wholesale markets, and impacts thereof, in different market settings, including general auction design in electricity wholesale markets~\cite{SpotMarketUK,ElecAuctions,IncCompMarketDesign_Low} with additional focus elements on the existence and properties of Nash equilibrial~\cite{SpotMarketUK,borenstein1997competitive} and the impacts of transmission capacity limitation in zonal and nodal markets~\cite{holmberg2015comparison,sarfati2020simulation}, among others. A large body of literature has also focused on the bidding behavior of aggregators in ancillary services markets~\cite{AggregatorFlexBidding}. Such strategic behavior can also apply to \ac{TSO}-\ac{DSO} coordinated flexibility markets.

As in any competitive electricity market, \acp{FSP} can be expected to bid strategically, choosing offers to maximize profit given their marginal costs, the market design (structure, pricing, and clearing rules), and beliefs about rivals’ bids. In imperfectly competitive settings, such behavior can materially affect market efficiency. Since incentives differ across different TSO-DSO coordinated flexibility market designs, evaluating bidding incentives in each is essential to determine: (i) which strategic behaviors a design induces, (ii) their efficiency impacts, (iii) the mechanisms driving them, and (iv) mitigation measures to limit efficiency losses. 

For instance, the works in \cite{Marques2023_Gaming,MARQUES2024_Gaming} have focused on the impact of \ac{FSP} strategic behavior on the efficiency of different \ac{TSO}-\ac{DSO} coordinated market models.\footnote{This approach has been evaluated as part of the CoordiNet~\cite{CoordiNetWebpage} and OneNet~\cite{OneNetWebpage} projects.}  
This requires first determining how \acp{FSP} optimally select their bids under different market structures, and subsequently evaluating market efficiency when these strategically chosen bids are submitted. 
As an \ac{FSP}’s revenue and bidding incentives depend on these interactions, strategic behavior can naturally be represented as a game among competing agents. 

The results in \cite{Marques2023_Gaming,MARQUES2024_Gaming} show that rational and strategically motivated \acp{FSP} have substantial opportunities to manipulate prices, and that the extent of this impact depends on the coordination scheme (common, fragmented, or multilevel), network topology, grid constraints, and the number and spatial distribution of \acp{FSP}. Fragmented market designs, in particular, increase the likelihood of market power as they reduce the volume of competing bids available to system operators, thereby weakening competition in each market stage. Unpriced interface flows further accentuate this effect by creating additional transmission-level needs that must be resolved using a smaller set of bids. In contrast, joint or common markets support higher liquidity and can partially mitigate the impacts of unilateral or collusive strategic behavior. Structural congestion can also give rise to local market power or even monopolistic conditions when parts of the network become electrically isolated and served by only a few \acp{FSP}, an issue that affects all \ac{TSO}-\ac{DSO} coordinated market models. These findings indicate that network constraints and the spatial concentration of \acp{FSP} must be actively monitored when designing coordinated flexibility markets.

Several works in the literature have in particular focused on the \ac{FSP} strategic behavior that can arise in sequential market structures~\cite{Sarfati2018}. A two-level game was developed in~\cite{Migliavacca_D23_2023} to compare a few \ac{TSO}-\ac{DSO} coordination schemes considering a sequence of \ac{DA} followed by the \ac{AS} market. The work explored different architectures: namely, (S1) a one-stage architecture in which wholesale energy and \ac{TSO}-\ac{DSO} services are jointly optimized, (S2) a two-stage structure in which the \ac{DA} market runs first followed by a joint \ac{TSO}-\ac{DSO} market, (S3) a three-stage setting in which the \ac{DA} market runs first followed by two separate \ac{TSO} and \ac{DSO} auctions in which each \ac{SO} can only access resources connected to its own network, and (S4)  a three-stage structure composed first of the \ac{DA} energy market, followed by a local \ac{DSO} market, whose unused flexibility is then forwarded to the third layer in which the \ac{TSO} procures flexibility from the residual distribution-level flexibility and transmission-level flexibility. The work identified a risk in S4 due to strategic behavior concerning load-shedding actions in which distributed flexibility can manipulate its bidding behavior, including the submission of elevated prices sure to be cleared at the transmission level to reap additional financial benefits. The work reverted to a preference towards S3 due to that risk.    

\subsection{Key Practical Challenges}\label{subsubsec:MarketChallenges} The body of literature on TSO-DSO coordinated flexibility markets have already introduced solid contributions, enabling initial piloting activities and moving those pilots towards operational/commercial implementation. There are, nonetheless, still a number of theoretical and practical challenges facing those markets, a key set of which is highlighted next.

\subsubsection{Network Representation: Market Clearing Constraints vs. Actual Grid Power Flows}

Different network formulations can be included in market-clearing problems. However, as these are optimization problems, using non-convex formulations (such as \ac{AC} power flows and 3-phase unbalanced representations for the distribution grid) can make the problem solution intractable\footnote{Other discrete variables pertaining to complex bid types or to discrete grid topology reconfiguration actions can even further complicate these problems and their decomposition possibilities.}. For these reasons, different market-clearing formulations opt for relaxed and/or linearized power flow formulations, such as linear or second-order cone programming formulations. However, these relaxations do not always naturally compute feasible solutions of the original problem, except for specific conditions, which can at instances be difficult to meet for more complex practical grid models. 
To avoid grid complexity, the market problem can be, for example, adapted to specific critical limitations of the grid to be considered, via a set of critical lines/elements, instead of a full network representation, supported by abstraction techniques such as via pre-qualification mechanisms or \ac{OE}-based methods. Indeed, the flexibility market-clearing goal is not to run precise power flow calculations, but rather to ensure that the flexibility procured effectively solves the flexibility needs of the system, without causing additional constraint violations. This, as a result, relaxes the need for precise power flow modeling. 

In addition, quite often, full visibility of the \ac{DN} is not available to the DSO~\cite{VANIN2023,Koirala2024}, which puts limitation on the complexity of the market clearing grid models that can be implemented. In this regard, stochastic representations can be considered to accommodate this uncertainty~\cite{delchambre2024phase, delchambre2025influence}.

\subsubsection{Baselining}

One key requirement for the use of distributed flexibility in a (TSO-DSO coordinated) flexibility market setting is the ability to verify the volume of flexibility actually delivered. This volume is measured as the change in net consumption over a certain period with respect to a counterfactual consumption level, i.e., what the consumption level would have been, had no flexibility been activated. This counterfactual consumption profile is known as the \emph{baseline}. Computing this baseline is quite challenging, especially when considering an aggregation of different types of flexibility assets~\cite{LIND2024_Baseline}. A number of baseline methods have been proposed in the literature, which rely on historical averages (known as X of Y methods, considering averaged consumption values from a subset of X days from a set of Y valid historical days), or direct metering before and after activation (known as meter-before, meter-after), and other machine-learning-based forecasting methods that factor in other impacting features such as weather data and location among others, to estimate baselines. The work in~\cite{LIND2024_Baseline} has surveyed the different proposed baseline methods in the literature and analyzed them with respect to the service to be delivered to the grid and the nature of the flexibility. Different methods return different scoring in terms of their accuracy, simplicity, and integrity (i.e., robustness to manipulation), hindering their practical implementation. Baselining is still to date one of the main challenges to be faced by flexibility markets which use low/medium-voltage flexibility.  

\subsubsection{Scalability}

Different TSO-DSO coordinated flexibility markets, especially when pooling a large number of flexibility bids which may also include binary/integer conditions on their purchasing, can face practical computational challenges. In addition, iterative approaches proposed to solve TSO-DSO coordinated markets are not always guaranteed to converge, and if they do, their convergence time can become challenging for practical implementation. This challenge is typically mitigated by simplifying assumptions on the constraints and processes included and the stochasticity thereof. As such, there is a need for the introduction of efficient solution mechanisms for these markets, when they enter their practical operational stage. 

For example, recent works \cite{MUsman2023} have focused on addressing more complex \ac{TSO}-\ac{DSO} coordination models, including essential yet challenging practical operation requirements, i.e. the N-1 security criterion at \ac{TSO} level, operation uncertainties and time coupling at both \ac{TSO} and \ac{DSO} levels, and emerging flexible assets (energy storage devices and flexible demand with intertemporal constraints). This leads to a set of large \ac{NLP} problems to be solved.

There is, generally, no approach that retains tractability in the presence of the most computationally heavy model features (stochasticity, multi-periods, N-1 security, and -- quite importantly -- realistic grid representation), thus requiring simplifications for practical implementation.  

In summary, the challenges faced in terms of network accuracy, scalability, and baselining are key research directions to be further pursued to fully support the practical implementation of TSO-DSO coordinated flexibility markets, and fully unlock their potential.   

\section{Technological Aspects}

\subsection{Architectures of OT and IT Systems in DNs to Enable Flexibility Provision}

\begin{figure}[h!]
    \centering
    \includegraphics[
        width=0.75\textwidth,
        trim=280 110 40 110,
        clip
    ]{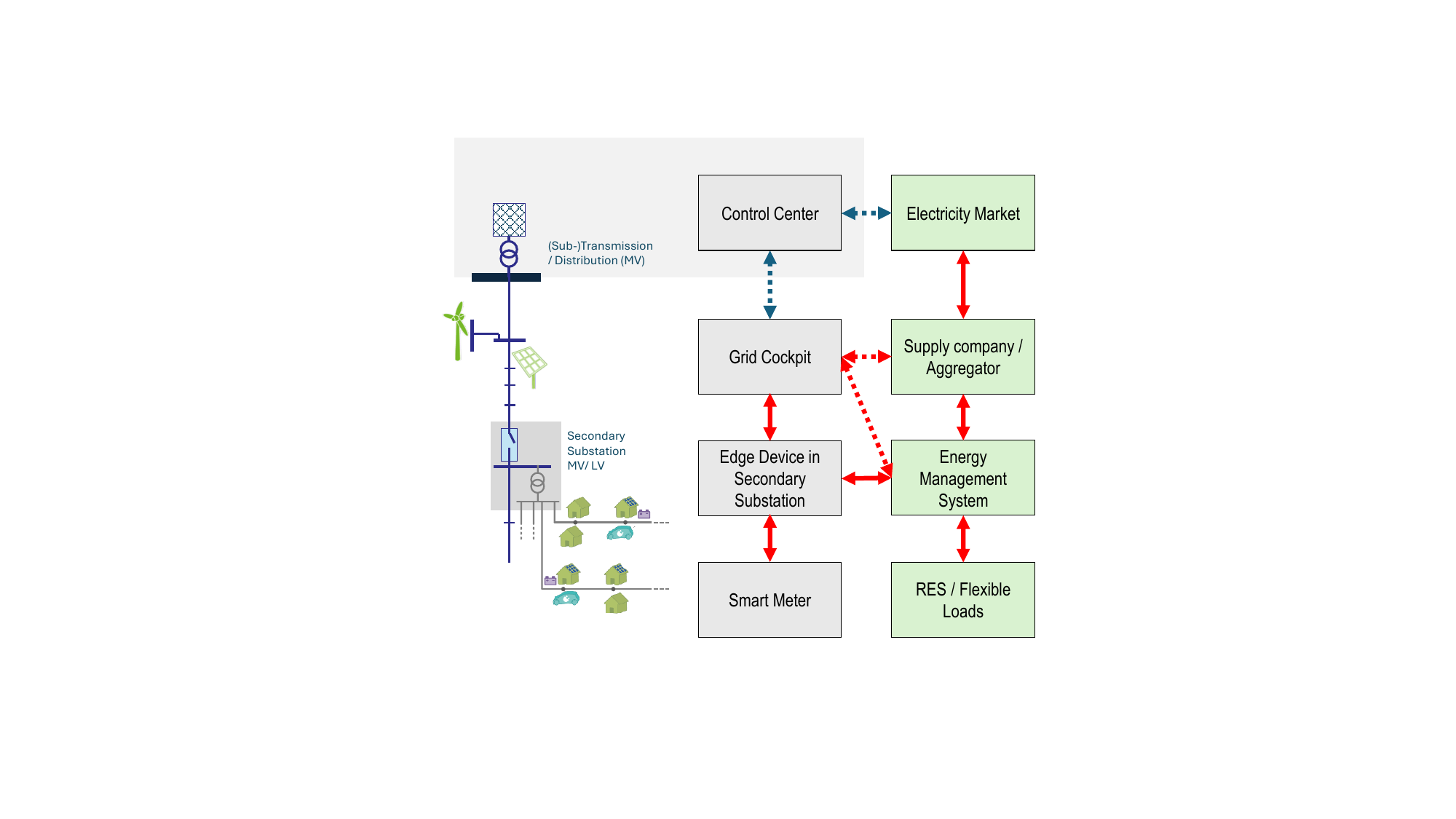}
    \caption{TSO-DSO interaction related OT and IT systems and major functions}
    \label{fig:tech_aspects}
\end{figure}

A clear distinction must be made between market processes and grid processes when analyzing the technical aspects of TSO–DSO interaction in unbundled power systems. In Fig.~\ref{fig:tech_aspects}, the right column represents the principal market‑related functions, ranging from wholesale electricity markets to supply companies (load serving entities) and aggregators, and extending to \ac{EMS} deployed in residential, commercial, and industrial facilities. \acp{RES} and flexible loads appear at the lower end of this column, forming the interface between local energy resources and market‑oriented coordination mechanisms. \ac{DERMS} may operate at the \ac{EMS} level for localized optimization or at the aggregation level for regional coordination. 

The left column of Fig.~\ref{fig:tech_aspects} contains the grid‑related systems. These include control centers—primarily deployed at transmission and subtransmission levels—and grid cockpits, which provide situational awareness and supervisory functions for \ac{MV} and \ac{LV} networks. Below these, intelligent edge devices have emerged as a new operational layer. These devices coordinate protection, automation, and monitoring functions within secondary substations, with their specific roles varying according to grid topology and national system architecture. In many European distribution systems, such devices are typically located at the interface between \ac{MV} and \ac{LV} levels. In North America, they are integrated into the \ac{MV} distribution grid. These edge devices can also serve to manage, control and protect large-scale wind, PV or even EV charging parks equally to secondary substations, but with adapted functions setups.

Smart meters constitute the lowest layer of the grid‑side architecture. While originally designed for billing purposes, they increasingly provide time‑synchronized measurements relevant to grid state estimation, power flow monitoring, and \ac{DG} visibility. However, this strongly depends on regulatory frames in the respective countries.

Although conceptually distinct, market processes and grid processes exhibit strong interdependencies, particularly in congestion management. Classical control centers have traditionally operated within isolated \ac{OT} environments based on dedicated hardware, proprietary communication protocols, and physically segregated networks. Despite this isolation, \ac{OT} systems require information from external \ac{IT}‑based systems and must also transmit operational requests to them. For example, control centers rely on generation schedules from market participants and must issue setpoint changes or redispatch instructions during congestion events. At lower grid levels, \acp{DSO} collect information on \ac{DG} and provide it to \acp{TSO} to support system‑wide operational planning. In several jurisdictions, redispatch signals for \ac{DG} units are transmitted from higher to lower grid levels, requiring coordinated \ac{OT}–\ac{IT} data exchange.
At transmission and subtransmission levels, \ac{SCADA} systems within the \ac{OT} domain remain the standard for real‑time control center operation. In contrast, \ac{MV} and \ac{LV} networks still lack widespread monitoring infrastructure, creating visibility gaps that must be addressed through emerging technologies such as edge devices and advanced metering. A structural distinction persists between \ac{OT} systems used for real‑time grid operation and \ac{IT} systems supporting market processes, customer services, and enterprise functions. 

\ac{IT} systems typically rely on public communication infrastructures, including the internet and mobile networks. Consequently, effective TSO–DSO interaction increasingly depends on secure, interoperable data exchange mechanisms bridging \ac{OT}‑ and \ac{IT}‑based environments.
A functional bifurcation is also emerging between \acp{DNO}, responsible for asset management, maintenance, outage response, and traditional reliability objectives (gray), and \acp{DSO}, responsible for \ac{DER} coordination, flexibility markets, real‑time optimization, and digitalization (green). In North American practice, Grid Cockpit functionalities are often embedded within \acp{ADMS}, which integrate \ac{OT}‑based monitoring with \ac{IT}‑based analytics and forecasting, often supported by the \ac{AMI}. Notably, a growing trend toward convergence of \ac{DNO} and \ac{DSO} roles is observable—particularly in \ac{DER}‑dominated and highly digitized grids—leading to unified \ac{DSO} models even within vertically integrated utilities. This evolution underscores the need for adaptive regulatory frameworks and technical architectures capable of supporting hybridized operational responsibilities.

The following subsections detail emerging grid supervision and control systems, including grid cockpits and edge devices, as well as the corresponding market‑side systems such as aggregators and energy management platforms. Trends in communication and data infrastructure are also examined, with emphasis on their implications for secure and scalable \ac{TSO}–\ac{DSO} coordination.

When prioritizing system functions, the most critical component is the grid cockpit at the distribution level, as it represents the digital counterpart to the \ac{TSO} control center. It serves as the central hub for aggregating all operationally relevant data. Edge devices can enhance system resiliency; however, certain functionalities may also be integrated directly into the grid cockpit.
On the market side, the greatest added value is provided by aggregators and energy management systems, which collect distributed flexibility and translate it into services for both the market and the grid.

\subsection{Grid Cockpit}

Highly branched and ramified \acp{DN} exceed the capabilities of current \ac{SCADA} architectures. In these network segments, cloud-based solutions from vendors are increasingly deployed \cite{CKoehler2024}. An open-source activity for flexible modularization of functions has been established \cite{AGoering2016}. These systems represent grid structures in a topological form and integrate technical data, including grid state information, for analytical and operational purposes. They rely on \ac{IT} and therefore use cloud infrastructures \cite{SongZhang2022} as well as public internet and mobile communication networks for the transmission of measurement data and control signals.
Key specifications of such grid cockpits include:

\begin{itemize}
  \item Adaptation to customer requirements solely through parameter settings
    \begin{itemize}
      \item only one software version is maintained for all customers
      \item no development of customer-specific software versions
      \item central patching of instances within a few minutes
    \end{itemize}
  \item Strict separation between core functionalities, modular applications, and customer-specific parameters such as grid data and measurements
      \begin{itemize}
      \item software applications and functions can be added when needed
      \item interfaces between \acp{TSO} and \acp{DSO} can be implemented with low effort, for example, for redispatch at \ac{DSO} level
    \end{itemize}
     \item Use of security mechanisms inherent to cloud technologies
\end{itemize}

Modern grid cockpits, which can safely be considered as specialized applications within \ac{ADMS}, aim to serve as the central digital twin for grid operation and, increasingly, for grid planning and asset management \cite{CKoehler2024}. As a result, they integrate a growing number of \ac{DSO} processes within one system architecture. Traditionally, control center operation, grid planning, and asset management have been organized as separate processes supported by distinct software tools and data models.

\subsection{Edge Devices}

An edge device in a substation is essentially a local system for computation and measurement \cite{Samie2019}. The concept is to deploy a single device that integrates multiple functions rather than several dedicated devices that each perform only one or a few functions, like a protection or fault direction indication device today. The number and content of the functions can be updated and extended over time \cite{Velasquez2019}. 

The fundamental role of an edge device is to acquire raw measurement data, pre-process it, and provide it to the local applications of the system. To enable the execution of several functions on one device, the individual applications must be encapsulated, for example, through software virtualization and container technologies \cite{Narayan2020}. At lower grid levels, which are not supervised manually through control centers, edge devices constitute the primary data sources for grid cockpits. They filter measurement data, detect malfunctions, supply topology information \cite{TSchwierz2022}, and forward the processed data to the grid cockpit. Local measurement units can achieve high sampling rates up to 200 kHz, depending on the hardware provider. This enables additional functionalities (besides protection and fault direction indication) such as power quality monitoring, state estimation \cite{DHilbrich2022}, phasor measurement \cite{Mohsenian-Rad2023}, congestion management, and coordination of local voltage control \cite{RPalaniappan2019}.  

Depending on the operational strategy of the \ac{DSO}, certain functions are executed directly on the edge device, whereas others are implemented centrally in the grid cockpit. Functions that rely on measurement data with a low update frequency, such as state estimation, can be implemented centrally but heavily rely on the communication infrastructure.
For improving system resiliency, it may be advantageous to execute specific functions locally on the edge device. For instance, congestion in \ac{LV} or \ac{MV} grids can be identified through a local state estimation and mitigated by temporarily limiting flexible power consumption on the customer side, such as \ac{EV} charging or \ac{RES} control \cite{Holt2020}. Such a function can solve problems automatically on the lowest grid levels. However, the grid data for such functions, which are usually maintained centrally in the grid cockpit, have to be provided as a set of parameters to the edge devices.  
Simulation plays a critical role in bridging observability gaps where \ac{SCADA}, \ac{AMI}, and grid‑edge sensing remain limited. Modern \ac{ADMS} platforms from major vendors increasingly rely on this capability to support real‑time state estimation in partially or fully unobservable feeders, generating hybrid measurement sets that combine real and pseudo‑measurements \cite{AjayYadav2023}. When data is scarce, these simulation‑enhanced estimates significantly strengthen the situational awareness and decision‑support functions of grid cockpits.

\subsection{Data}
Effective coordination across grid levels requires the exchange of information while maintaining the confidentiality of certain data. This applies both to interactions between grid operators and to relationships among market participants. For this reason, decentralized and distributed database concepts are increasingly being considered.
Decentralized databases become part of a federated system through a set of essential services and agreements that enable their interoperability.

Federation is widely regarded as a foundational principle for achieving sovereign data exchange among participants on a peer-to-peer basis \cite{Pourebrahimi2005}. Such systems reconcile two competing requirements by allowing an appropriate level of information sharing while preserving the autonomy of each participating component, in this case, the individual databases \cite{Heimbigner1985}. Federated systems typically exhibit physical distribution, heterogeneity, and varying degrees of operational independence \cite{Busse2020}. Overall, this approach reduces the amount of data that must be exchanged between system instances.
Effective TSO–DSO coordination increasingly depends on high-quality, high-frequency data exchange supported by cloud storage and \ac{AI} analytics \cite{SongZhang2025}.

As grids generate massive volumes of \ac{SCADA}, \ac{AMI}, \ac{PMU}, \ac{DER}, and grid-edge data, cloud platforms provide the scalable infrastructure to aggregate and standardize these datasets into a shared operational layer. \ac{AI} models trained on this unified data improve load and generation forecasts, detect anomalies, and simulate congestion or flexibility scenarios across both transmission and distribution boundaries. This continuous, data-driven exchange replaces periodic reporting with real-time situational awareness, enabling \acp{TSO} and \acp{DSO} to jointly manage variability, congestion, and \ac{DER} flexibility with greater precision and transparency.

\subsection{Communication}

The underlying communication paradigm is that energy, power, and communication systems are mutually dependent and together form cyber-physical energy and power systems \cite{YCao2019}. In this context, an important paradigm shift is the move from “security by obstruction” to “security by design.” Rather than treating cybersecurity as a mechanism that obstructs, restricts, or isolates communication, security by design embeds cybersecurity requirements into the architecture of cyber-physical energy and power systems from the outset. Cybersecurity then acts as an enabler that protects the integrity, availability, and confidentiality of the information flows required for innovative approaches, including monitoring, coordination, and restoration. Accordingly, the communication system must remain capable of delivering its services during and after major disturbances, including blackouts and cyberattacks. It should operate on a regional basis and must not exhibit single points of failure that could trigger system-wide outages, as observed, for instance, when the failure of routing servers in mobile communication networks led to a complete loss of service \cite{StefanMonhof2018}.

Local and regional communication infrastructures should therefore be designed to be resilient against disturbances, enabling the coordination of lower grid levels and, in the future, the operation of islanded regions for emergency supply \cite{NilsDorsch2018}. The next generation of 6G mobile communication should be designed to fulfill the criteria for critical infrastructures, such as distributed power and energy systems \cite{Mahmood2021}.

Communication at the edge is becoming a cornerstone of TSO–DSO coordination, supported by a telecom stack that spans \ac{NAN}, \ac{WAN}, and substation networks. At the distribution level, edge computing devices rely on \ac{NAN} technologies—RF mesh, \ac{PLC}, Wi‑Fi, LTE/5G, and MQTT/AMQP messaging—to collect, fuse, and validate high‑frequency data from sensors, \acp{DER}, and protection devices \cite{JStoupis2023}. These \ac{NAN} infrastructures must meet strict latency, reliability, and interoperability requirements, following standardized protocol stacks such as DNP3, Modbus TCP, IEC 61850 mappings, and OSI/TCP‑IP layering \cite{BGoswami2024}. Upstream, \ac{TSO}‑level substations increasingly operate on IEC 61850‑based Ethernet networks enhanced by software-defined networking and programmable data planes, enabling deterministic routing, multicast management for GOOSE/SV traffic, and in‑band telemetry for real‑time visibility \cite{SGutierrez2023}. Together, this integrated telecom‑protocol fabric—edge‑level \ac{NAN} connectivity, standardized data models, and programmable substation networks—creates a synchronized, low‑latency communication backbone that allows \acp{TSO} and \acp{DSO} to exchange high‑quality operational data and coordinate actions in real time.

\ac{IT} systems that rely on cloud infrastructure and the internet are highly centralized. In the future, an alternative may be decentralized distributed systems equipped with edge functionalities. Such systems could operate in island mode, thereby offering high resilience to the overall power system. Decentralized functions would monitor, for instance, individual branches of the \ac{LV} network.
Considering this architecture more broadly, the interactions across all grid levels, not just the interaction between \acp{TSO} and \acp{DSO}, impose numerous requirements when providing services such as power flexibility for grid and market use. These requirements include distributed communication, data, and computing infrastructures at each grid level and within each region. The resulting structure forms a system of systems, referred to as a \textit{holonic system} architecture in \cite{ChrisRehtanz2024}, which specifies the details of such an approach.

\subsection{Aggregator / Energy Management System}

Aggregators and \ac{EMS}, including \ac{DERMS}, form the market-side counterpart of the grid-side digital platforms described above. Together, they provide the mechanisms through which \acp{DER} become visible, controllable, and dispatchable for both \acp{TSO} and \acp{DSO}~\cite{Sugunaraj2026}. Their role is essential in unbundled and \ac{DER}-dominated power systems, where flexibility must be mobilized across millions of small-scale devices while ensuring \ac{DN} feasibility.
At the customer level, household and industrial \ac{EMS} coordinate devices such as \acp{HP}, \acp{PV}, \ac{BESS} systems, \acp{EV}, and other flexible loads. They respond either to local optimization objectives—such as self-consumption and time-of-use tariffs—or to external price or flexibility signals supplied by an aggregator or retailer. Through such schemes, local flexibility is extracted and transformed into standardized services that can be offered to electricity markets or to grid operators \cite{Papalexopoulos2024}. Depending on local regulation, the connection capacity may be fixed contractually or dynamically adjusted by the grid operator during congestion events via setpoints originating from the grid cockpit or edge devices.

\begin{figure}
    \centering
    \includegraphics[width=0.45\textwidth]{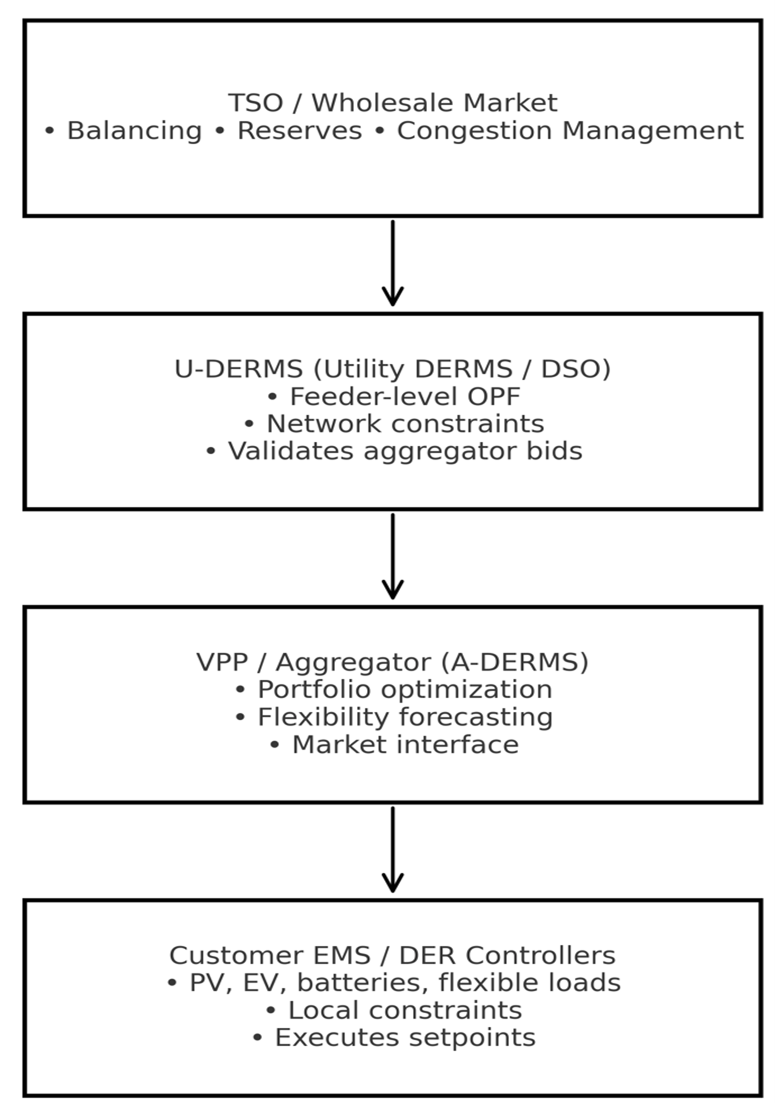}
    \caption{Technologies for hierarchical coordination among TSO/wholesale market: U-DERMS (DSO), VPP/A-DERMS, and customer EMS/DER controllers.}
    \label{fig:tech_hierarchy}
\end{figure}

At a higher level, aggregators combine the flexibility of large numbers of customers and participate in wholesale electricity markets, balancing markets, or \acp{AS}. They also provide flexibility directly to \acp{DSO} for congestion management, voltage support, or remedial action schemes. In this respect, aggregators act as intermediate system orchestrators, translating device-level behavior into aggregated flexibility envelopes while ensuring compliance with \ac{DSO}-imposed constraints and \ac{TSO} scheduling requirements. This aligns with the evolution from traditional demand response schemes to fully integrated \acp{VPP}, which operate as software-defined power stations \cite{InnoKamwa2025} capable of delivering voltage and frequency response, fast reserve, congestion management, and even black-start–like functionality across diverse jurisdictions.
Recent research on \ac{DER} coordination proposes a unified hierarchical \ac{DERMS} architecture that naturally complements \ac{TSO}–\ac{DSO} operational coordination \cite{PPGavgani2025}. Three layers, depicted in Fig.~\ref{fig:tech_hierarchy} can be distinguished:
\begin{enumerate}
    \item \ac{U-DERMS} — Operated by the \ac{DSO}, with access to the full \ac{DN} model, state estimation functions, and operational constraints. The \ac{U-DERMS} validates all flexibility bids submitted by aggregators, ensures \ac{DN} feasibility, and performs constrained optimization to coordinate \ac{DER} flexibility with grid-side objectives.

    \item \ac{A-DERMS} — Operated by flexibility service providers, responsible for managing thousands of heterogeneous \acp{DER}, forecasting available flexibility, and constructing feasible flexibility regions that can be communicated to the \ac{DSO} and \ac{TSO}. The \ac{A-DERMS} functions as an abstraction layer, transforming individual \ac{DER} constraints into aggregated models suitable for network-level optimization.
    \item R-Level Controllers — Located in customer \ac{EMS} or device-level controllers, execute setpoints derived from \ac{A-DERMS} scheduling. These controllers ensure local feasibility while obeying distribution-level constraints passed down through the hierarchy.
\end{enumerate}
%
%
This hierarchical design appears consistently across three complementary studies \cite{PPGavgani2025,PourGavgani2024,PouyaGavgani2025}. The first formalizes a \ac{DERMS} hierarchy capable of managing heterogeneous \ac{DER} fleets at scale. The second introduces mathematically grounded aggregation techniques—such as homothetic polytopes, voltage-sensitivity estimation, and the Sequential Adjustment Method—allowing the \ac{U-DERMS} to solve reduced-complexity \ac{OPF} problems with reliable real-time voltage compliance. The third integrates \ac{VPP} concepts into the \ac{DERMS} stack, highlighting that future grid flexibility will not be delivered by individual devices but by aggregated, software-defined portfolios with predictable behavior and standardized interfaces. In this framework, \acp{VPP} and \ac{A-DERMS} become operational intermediaries that translate \ac{TSO} requirements—balancing needs, congestion limits, remedial actions—into feasible feeder-level setpoints issued through \ac{U-DERMS}. Conversely, the \ac{DSO} validates and constrains aggregator bids before allowing participation in wholesale markets, ensuring that system-level actions remain consistent with local grid conditions. This two-way dependency reinforces the central role of aggregators in future TSO–DSO interoperability structures.
Overall, aggregators, \ac{EMS}, \ac{DERMS}, and \acp{VPP} constitute the “market-side digital ecosystem” that mirrors the grid-side \ac{OT}/\ac{IT} stack (control centers, \ac{ADMS}, grid cockpits, and edge devices). Together, they establish the cyber-physical coordination fabric required for next-generation flexibility mobilization, enabling \acp{DSO} and \ac{TSO} to activate \ac{DER}-based services safely, efficiently, and at scale.

All coordination and interaction between different systems on both the grid and market sides must account for computation and communication latencies in order to provide specific services. In particular, ancillary services for grid stability may require fast and strictly real-time behavior.
Control center actions initiated by the \ac{TSO} and propagated to the \ac{DSO} for congestion management during thermal overload can typically be executed within several minutes and still remain effective. In contrast, voltage stability issues following faults must be resolved within seconds, which can impose time-critical constraints on cascaded system architectures.

Market-related services generally operate on slower timescales—typically no faster than 5 minutes before the 15-minute market interval—and are therefore largely uncritical from a real-time perspective across different system instances. Fast ancillary services, however, may need to be prepared and coordinated preventively at higher system levels, such as the control center or the grid cockpit, while their execution must occur locally at the lowest feasible level, for instance, within edge devices or energy management systems during event-driven situations. Overall, coordination between the different systems remains a challenge, despite advances in computer systems, and requires careful investigation.

\section{Conclusions and Outlook}

This paper has highlighted the complexities of TSO–DSO coordination and introduced some challenges that remain unaddressed or not fully solved.

\subsection{On Flexibility Harvesting and DN Equivalents}

Aggregators and \acp{VPP} are expected to manage thousands of devices for service provision while respecting device owners' objectives. Although several works have focused on aggregation of the same device type, e.g., only water heaters or only \acp{EV}, coordinating heterogeneous \acp{DER} might unlock greater flexibility. Additionally, scenarios involving multiple aggregators operating within the same \ac{DN} need to be further analyzed to identify potential counterproductive effects on the local grid arising from different service provisions. 

Novel strategies are needed to increase flexibility in grid-constrained areas, as locational network constraints may reduce \acp{DER} and flexible load participation in grid services, driven by feeder conditions rather than device capabilities. This could create disparities, as some end-users would participate consistently in service provision, while others would be frequently curtailed by grid operational constraints. 

Future studies attempting to aggregate small-scale devices to support \ac{TN}'s voltage and frequency need to account for the potential disconnection of devices during large disturbances, and investigate how long \ac{DER} and flexible loads can sustain a service. Also, the \acp{FR} are mainly computed for normal operating conditions. The extent to which these regions shrink during \ac{TN} disturbances needs to be evaluated.   

Recent events such as the April 28, 2025 Iberian Peninsula blackout illustrate how inadequate voltage and reactive-power control, together with generation disconnections, can contribute to cascading failures \cite{sabolic2025april}. Therefore, simplified P/Q representations may be insufficient when assessing voltage-support flexibility. Where possible and applicable, \acp{DER} should be modeled using voltage-control-capable representations, such as PV-bus representation in static analyses and explicit voltage-control models in dynamic studies.

In terms of \ac{ADN} equivalents for \ac{TN} studies, the proposed \textit{black-box} models depend heavily on the quality of the training data. Future \textit{gray-box} models should better account for voltage diversity within the \ac{DN} and consider the variation in load and \ac{DG} composition across different times of day and seasons. Hence, robustness to uncertainty and limited information is needed to increase the trust in these models. Finally, more work is needed to guide \acp{DSO} on the information they must provide to the \ac{TSO} to build these dynamic equivalents.   

\subsection{On Coordination Mechanisms}

While mathematically appealing, mainly due to non-convexity challenges (e.g., intrinsic non-convexity of equations, potential region holes and discontinuities), the topic of computing approximately the boundary of \ac{ADN} feasible region, in the space of active/reactive powers at the interface with the transmission system, has remained a favorite research topic. However, the paper argued that this topic has been conceptually and computationally narrowly approached via over-simplified methods. More importantly, this approach is inadequate for TSO-DSO coordination because it omits the indispensable utilization requirement of the cost of shifting active/reactive power within the region. Finally, the non-convexity challenges are also common to many other problems in power systems space; focusing further research in this topic may not solve yet further complicates an already very complex problem of TSO-DSO coordination while, in real-life, practical approaches are sought. 

This paper has strongly advocated superseding such active/reactive power flexibility regions by simpler and practical ranges of active/reactive power flexibility of \ac{ADN} at the interface with the transmission system. Such an approach reflects practical characteristics of power systems (strong coupling between current and active powers as well as voltage and reactive powers) and fits the current structure of markets for \acp{AS} (e.g., congestion management, voltage control). As argued, the most important benefit of adopting flexibility ranges is to enable computing approximate (e.g., piece-wise linear) cost curves of active and reactive power flexibility. Such cost curves could be seamlessly integrated, without compromising computation efficiency, in \ac{TSO} tools for procuring at the least cost flexibility, including from \acp{ADN}. Still, computing tractably and with mathematical guarantees of feasibility such cost curves, especially under uncertainties and multiple coupled time periods, is an open question, for which reasonably accurate and scalable approximations are to be found. 
Despite the large number of papers proposing mechanisms to coordinate \ac{TSO} and \ac{DSO} operations, three major research gaps persist: 
\begin{itemize}
\item in terms of models employed, most works address over-simplified, inaccurate problems at both \ac{TSO} and DSO levels. Exact models should include key practical operation requirements, such as the N-1 security criterion at \ac{TSO} level, operation uncertainties and time coupling at both \ac{TSO} and \ac{DSO} levels, and emerging flexible assets (battery storage and flexible demand). Furthermore, these papers consider, at best, a few TSO-DSO interfaces, while in practice, there may be hundreds of such substations. Finally, the downward \acp{ADN} at some substations, depending on the voltage levels, may be vast and meshed, as opposed to small and radial, as assumed exclusively. Meshed \acp{DN} entail that the flexibility that could be provided across some substations is strongly coupled. Finally, there are limits on how much flexibility could be harvested bottom-up across voltage levels (from low to high) to support the \ac{TSO}, a topic that remains unaddressed. 
\item no proposed approach retains tractability in the presence of the most computationally heavy model features (stochasticity, multi-periods, N-1 security). The sizes of transmission and distribution systems on which coordination mechanisms have been tested are much smaller than in practice. 
\item Very few works consider the cost of aggregated active/reactive power flexibility of an \ac{ADN} at the TSO-DSO interface. 
\end{itemize}

In summary, despite immense research devoted to TSO-DSO coordination mechanisms, the lack of accuracy and (numerically demonstrated) tractability/scalability of the proposed methodologies prevents their practical adoption. Hence, from a practical perspective, similarly with many other topics tackled in academic power systems research, the work is still in its infancy. 

\subsection{On Algorithmic and Regulatory Tools}

From the algorithmic point of view, partitioning the social-welfare optimization of huge systems into distinct, yet mutually coordinated systems, reduces the computational complexity (as the number of decision variables is lower) but introduces the necessity to iterate between the solution of the different interconnected systems or between them and a coordination level (e.g. Benders’ decomposition) with possible consequences on convergence speed and even, sometimes, on the real attainment of a system optimum. On the regulatory ground, coordinating different entities that exchange only border information facilitates several aspects: there is no need to involve all decision-maker subjects at the same time, but each one optimizes its own subsystem, and data privacy constraints are fully respected because information on the single bidder does not go beyond the perimeter of the \ac{SO} responsible for its area. 

The need to replace system integration with system coordination will become even more significant in the future,  when the need will arise to coordinate together with the electric system all non-electric energy carriers (gas, hydrogen, district heating) in a holistic multi-energy prospect in order to facilitate the provision of services across the systems: in that case, not only the number of decision variables and decision makers will grow significantly, but, as the regulation of the different sectors is much different, the dialogue between the different subjects will become more complicated. In conclusion, the frontier, and the open questions for future investigation, consist in finding algorithmic and regulatory tools to cope with an increasing system complexity, which will push more in the direction of cooperation between the decision- maker subjects. In that sense, increasing the power of calculation machines will do just half the job. The rest will focus on deploying more powerful algorithms to solve coordinated systems, alongside regulatory measures to ensure effective coordination across separate actors, while maintaining a level playing field for all market participants and avoiding asymmetries that could enable discrimination.

\subsection{On the Technological Aspects}

The technical implementation of market services and \acp{AS} requires a joint consideration of hardware, software, and algorithms in the power system sector, as well as the associated data and communication systems. Manufacturers, however, face the challenge of achieving global standardization of technical solutions and processes while still accommodating the diverse regulatory market regimes. This is demanding because fundamental technical solutions must be adapted to the regional specificities of individual countries. Consequently, efforts must extend beyond purely technical aspects toward harmonizing the structures of market and system services. At the same time, any technical standardization must account for international developments arising from the rapid evolution of \ac{ICT} and data systems. Standards originating solely in the field of energy engineering may, in some cases, prove counterproductive.

\bibliographystyle{IEEEtran}
\bibliography{refs}

\end{document}